\documentclass[11pt,showpacs,preprintnumbers,superscriptaddress,amsmath,amssymb,nofootinbib,aps]{revtex4}
\usepackage{graphicx}
\usepackage{dcolumn}
\usepackage{bm}
\usepackage{amssymb}
\usepackage{amsmath}
\usepackage{epsfig}
\usepackage{color}
\usepackage{slashed}
\usepackage{hhline}
\usepackage{ulem}
\usepackage{bbm}

\def\be{\begin{equation}}
\def\ee{\end{equation}}
\newcommand{\bea}{\begin{eqnarray}}
\newcommand{\eea}{\end{eqnarray}}

\begin{document}


\title{Three-loop induced neutrino mass model in a non-invertible symmetry}

\author{Hiroshi Okada}
\email{hiroshi3okada@htu.edu.cn}
\affiliation{Department of Physics, Henan Normal University, Xinxiang 453007, China}

\author{Yoshihiro Shigekami}
\email{shigekami@htu.edu.cn}
\affiliation{Department of Physics, Henan Normal University, Xinxiang 453007, China}

\date{\today}

\begin{abstract}
We propose a new type of radiatively induced neutrino masses at three-loop level based on the Ma model, introducing a non-invertible symmetry in the class under a ${\mathbb Z_2}$ gauging of ${\mathbb Z_6}$ symmetry and adding three isospin doublet vector-like fermions $L'$ and singlet boson $S_0$. 
Under this symmetry, the Yukawa interactions directly related to the neutrino masses are not allowed at tree-level. 
However it is allowed at one-loop level due to $L'$ and $S_0$ as well as $\eta$, which is no longer invariant under this symmetry. 
Therefore, the symmetry is dynamically broken. 
Intriguingly, $\eta$ plays important roles in contributing to both the radiative matrices $y^\eta$ and $m_\nu$. 
After constructing our model, we show some numerical analyses to satisfy the lepton flavor violations, muon anomalous magnetic dipole moment, and a boson dark matter candidate $S_0$ or $\eta_R$ for the cases of normal hierarchy and inverted hierarchy. 
Then, we demonstrate allowed space for our input parameters. 
\end{abstract}
\maketitle
\newpage

\section{Introduction}
\label{sec:intro}

Non-invertible symmetries possess an intriguing nature that the symmetry can dynamically be broken at loop levels even though it is symmetric under the tree level.
Also, they have a kind of group theory such as ${\mathbb Z_N}$ symmetry. 
Thus, several literatures~\cite{Choi:2022jqy,Cordova:2022fhg,Cordova:2022ieu,Cordova:2024ypu,Kobayashi:2024cvp,Kobayashi:2024yqq,Kobayashi:2025znw,Suzuki:2025oov,Liang:2025dkm,Kobayashi:2025ldi,Kobayashi:2025lar,Kobayashi:2025cwx,Nomura:2025sod,Dong:2025jra,Nomura:2025yoa,Chen:2025awz} are recently appeared, applying these symmetries to phenomenology such as radiatively induced mass matrices, zero texture's realizations of mass matrices, mass eigenvalues and mixing angles, CP phases in quark and lepton sectors.
Although these symmetries are not groups, they are supported by some formal theories; the worldsheet theory of perturbative string theory \cite{Bhardwaj:2017xup}, heterotic string theory on toroidal orbifolds \cite{Dijkgraaf:1987vp,Kobayashi:2004ya, Kobayashi:2006wq, Beye:2014nxa, Thorngren:2021yso, Heckman:2024obe, Kaidi:2024wio}, Calabi-Yau threefolds \cite{Dong:2025pah}, and type II intersecting/magnetized D-brane models \cite{Kobayashi:2024yqq,Funakoshi:2024uvy}.

In this work, we apply a non-invertible symmetry in a class under ${\mathbb Z_2}$ gauging of ${\mathbb Z_6}$ symmetry to a radiative seesaw model, which is sometimes called Ma model~\cite{Ma:2006km}. 
The Yukawa coupling to generate the neutrino mass matrix is forbidden at the tree level due to this invertible symmetry, but it is induced at one-loop level. 

It would be helpful to start a brief review on the Ma model before discussing our model. 
The model provides the neutrino mass matrix at one-loop level, introducing the right-handed neutral fermions $N_R$ and an isospin doublet inert boson $\eta(\equiv [\eta^+,(\eta_R+i\eta_I)/\sqrt2]^T)$.
In addition, $\mathbb Z_2$ symmetry is imposed in order to forbid the tree-level neutrino mass matrix. 
The field contents and their charge assignments are listed in Table~\ref{tab:ma}.
Clearly, only new fields are assigned by $\mathbb Z_2$ odd.
Under the symmetry, the valid terms to induce the neutrino mass matrix as well as the charged-lepton masses are found as
\begin{align}
y^\ell_{i}{\overline L_{L_i}} H \ell_{R_i} + y^{\eta}_{ia} {\overline L_{L_i}} \tilde\eta N_{R_a} + M_{N_{a}} {\overline N^C_{R_a}}N_{R_a} 
+\lambda''_{H\eta} (H^\dag \eta)^2
+{\rm h.c.}, \label{eq:ma}
\end{align}
where $y^\ell$ and $M_N$ can be diagonal without loss of generality, and $\tilde\eta\equiv i\sigma_2\eta^*$ being $\sigma_2$ second Pauli matrix.
The resultant neutrino mass matrix is then given by 
\begin{align}
(m_\nu)_{ij} =\sum_{a=1}^3
\frac{ y^{\eta}_{ia} M_{N_{a}} y^{\eta}_{ja}}{(4\pi)^2}
\left[
\frac{m_R^2}{m_R^2 - M_{N_{a}}^2 }\ln\left(\frac{m^2_R}{M_{N_{a}}^2}\right)
-
\frac{m_I^2}{m_I^2 - M_{N_{a}}^2 }\ln\left(\frac{m^2_I}{M_{N_{a}}^2}\right)
\right]\equiv
y^{\eta}_{ia} D_{N_{a}} y^{\eta}_{ja}
,
\end{align}
where $m_R$ and $m_I$ are respectively mass eigenvalues of $\eta_R$ and $\eta_I$.
Note that the mass difference between $\eta_R$ and $\eta_I$ leads us to the non-vanishing neutrino masses and its difference arises from $\lambda''_{H\eta} $.
The neutrino mass matrix is diagonalized by a unitary matrix $U_\nu$ as $D_\nu\equiv U_\nu^T m_\nu U_\nu$.
Moreover, since the mass matrix of charged-lepton is diagonal, $U_\nu$ is identified by the lepton mixing matrix $U {= U_{\rm PMNS}}$ which could be observed. 
\begin{table}[!t]
\begin{tabular}{|c||c|c|c||c|c|}\hline\hline 
& ~$L_L$~ & ~$\ell_R$~ & ~${N_R}$~ & ~$H$~ & ~{$\eta$}~ \\\hline\hline 
$SU(2)_L$ & $\bm{2}$ & $\bm{1}$ & $\bm{1}$ & $\bm{2}$ & $\bm{2}$ \\\hline 
$U(1)_Y$ & $-\frac12$ & $-1$ & $0$ & $\frac12$ & $\frac12$ \\\hline
${\mathbb Z_2}$ & $+$ & $+$ & $-$ & $+$ & $-$ \\\hline
 \end{tabular}
\caption{The Ma model; charge assignments of the fermions and bosons
under $SU(2)_L\otimes U(1)_Y \otimes {\mathbb Z_2}$, where we assume all fermions have three families.}\label{tab:ma}
\end{table}
Following the Casas-Ibarra parametrization~\cite{Casas:2001sr}, we can rewrite $y^\eta$ in terms of observed and some parameters as
\begin{align} 
y^\eta = U^* \sqrt{D_\nu} O_N \left(\sqrt{D_N}\right)^{-1}
, \label{eq:y_eta}
\end{align}
where the $y^\eta$ has to satisfy the perturbative limit Max$[|y^\eta|] \lesssim 4\pi$, and $O_N$ is an orthogonal mixing matrix $O_N O_N^T =O_N^T O_N ={\mathbb I}$ with three complex free parameters.
The neutrino mass eigenvalues are written in terms of two observables ($\Delta m^2_{\rm atm} , \Delta m^2_{\rm sol}$) and the lightest neutrino mass eigenvalue as follows:
\begin{align}
{\rm NH}&:\ D_{\nu_2}= \sqrt{\Delta m^2_{\rm sol} + D_{\nu_1}^2},\quad D_{\nu_3}= \sqrt{\Delta m^2_{\rm atm} + D_{\nu_1}^2}, \label{eq:nh_neutmass}\\
{\rm IH}&:\ D_{\nu_1}= \sqrt{\Delta m^2_{\rm atm} - \Delta m^2_{\rm sol} + D_{\nu_3}^2},\quad D_{\nu_2}= \sqrt{\Delta m^2_{\rm atm} + D_{\nu_3}^2}\label{eq:ih_neutmass} ,
\end{align}
where NH (IH) is an abbreviated notation of normal hierarchy (inverted hierarchy). 

There are several experimental constraints on the model. 
First, the sum of neutrino masses, which is denoted by $\sum D_\nu=D_{\nu_1} + D_{\nu_2}+ D_{\nu_3}$, is constrained by the minimal standard cosmological model with CMB data as $\sum D_{\nu}\le$ 120 meV~\cite{Planck:2018vyg}. 
Second, the effective mass for neutrinoless double beta decay $m_{ee}$ is defined as 
\begin{align}
 m_{ee}= \left| D_{\nu_1} c^2_{12} c^2_{13}+ D_{\nu_2} s^2_{12} c^2_{13}e^{i\alpha_{21}}+ D_{\nu_3} s^2_{13}e^{i(\alpha_{31}-2\delta_{CP})} \right|,
\end{align}
where $s_{12,23,13} (c_{12,23,13})$, which are short-hand notations $\sin\theta_{12,23,13} (\cos\theta_{12,23,13})$, are neutrino mixing of $U$, $\delta_{CP}$ is Dirac phase, $\alpha_{21,31}$ are Majorana phases. 
The upper bound on $m_{ee}$ is provided by the current KamLAND-Zen data measured in the future~\cite{KamLAND-Zen:2024eml};
$ m_{ee} <(36-156)$ meV at 90 \% confidence level (CL). 
Moreover,$m_{\nu e}$ is given by
\begin{align}
m_{\nu e}=\sqrt{D_{\nu_1}^2 c^2_{13} c^2_{12} + D_{\nu_2}^2 c^2_{13} s^2_{12} + D_{\nu_3}^2 s^2_{13}}.
\end{align}
whose upper bound is given by KATRIN~\cite{KATRIN:2024cdt}; $m_{\nu e} \le 450$ meV at 90\% CL, which is weaker than $\sum D_\nu$ and $m_{ee}$.
In addition to constraints on the neutrino sector, we should care about those on the charged lepton sector in this kind of models, induced by $y^\eta$. 
The crucial constraints are coming from lepton flavor violation (LFV) processes and the muon anomalous magnetic moment ($g-2$), and their formulae are found as
\begin{align}
{\rm BR}(\ell_i\to\ell_j\gamma) &\approx \frac{48\pi^3 \alpha_{em} C_{ij}} {G_F^2 m_{\ell_i}^2}
\left( |(a_R)_{ij}|^2+|(a_L)_{ij}|^2\right), \label{eq:LFVs}\\
\Delta a_\mu &\approx - m_{\mu} \left[ (a_R)_{22}+(a_L)_{22} \right], \label{eq:mug-2}
\end{align}
where $\ell_1\equiv e$, ${\ell_2}\equiv \mu$, and ${\ell_3}\equiv \tau$, $\alpha_{em}$ is fine-structure constant, $G_F$ is Fermi constant, $C_{21}=1$, $C_{31}=0.1784$, and $C_{21}=0.173648$. 
$a_R$ and $a_L$ are given by
\begin{align}
(a_R)_{ij} &\approx \sum_{a=1}^3\frac{y^\eta_{ja} (y^\eta)^\dag_{ai} m_{\ell_i}}{12(4\pi)^2} 
\left[\frac{M_{N_a}^6 - 6 M_{N_a}^4 m^2_\eta + 3M_{N_a}^2 m^4_\eta +2 m^6_\eta+6 M_{N_a}^2 m^4_\eta\ln\left(\frac{M_{N_a}^2}{m^2_\eta}\right)}
{(M_{N_a}^2 - m^2_\eta)^4}\right], \label{eq:aR}\\
(a_L)_{ij} &\approx \sum_{a=1}^3\frac{y^\eta_{ja} (y^\eta)^\dag_{ai} m_{\ell_j}}{12(4\pi)^2} 
\left[\frac{M_{N_a}^6 - 6 M_{N_a}^4 m^2_\eta + 3M_{N_a}^2 m^4_\eta +2 m^6_\eta+6 M_{N_a}^2 m^4_\eta\ln\left(\frac{M_{N_a}^2}{m^2_\eta}\right)}
{(M_{N_a}^2 - m^2_\eta)^4}\right],\label{eq:aL}
\end{align}
where $m_\eta$ is the mass eigenvalue of $\eta^\pm$. 
The experimental upper bounds are respectively given by
\begin{align}
{\rm BR}(\mu\to e\gamma)\lesssim 3.1\times10^{-13},\quad
{\rm BR}(\tau\to e\gamma)\lesssim 3.3\times10^{-8},\quad
{\rm BR}(\tau\to \mu\gamma)\lesssim 4.4\times10^{-8}.
\end{align}
Recent measurement of muon $g-2$ includes the prediction of the Standard Model (SM)~\cite{Muong-2:2025xyk,Aliberti:2025beg}, and it is given within 1$\sigma$ as
\begin{align}
\Delta a_\mu \simeq (39\pm 64)\times 10^{-1}.
\label{eq:Delamu}
\end{align}
Now that experimental value allows both the sign of muon $g-2$, and the theory suggests negative. 
Therefore, it would be verifiable in the near future. 

It is notable that there are two dark matter (DM) candidates in the model: $N_{R_1}$ or $\eta_R$.
If the DM is fermion $N_{R_1}$, the main annihilation to explain the relic density of DM is arisen from Yukawa coupling $y^\eta$~\cite{Kubo:2006yx},
and only when $y^\eta={\cal O}(1)$, it reaches the observed relic density~\cite{Planck:2015fie}. 
Note that since $N_{R_1}$ does not couple to quark sector at the tree-level, one can easily evade constraints of direct detection searches.
In the case where the DM is $\eta_R$, the main cross section would come from kinetic terms~\cite{Hambye:2009pw}~\footnote{ If $y^\eta$ is almost perturbative limit or the DM mass is localized at nearby the resonance of half of the SM Higgs mass, the relic density would be explained by itself.}, which results in mostly fixed DM mass as
\begin{align}
m_R\approx 534 \pm 8.5 \ {\rm GeV}\ ({\rm 1\sigma}).
\end{align}
Although $\eta$ interacts with quark sector via $Z$-boson or Higgs boson at the tree level, it is easy to evade the constraints of direct detections. 
In case of interaction with $Z$-boson, which is inelastic scattering, the constraint would be evaded when the mass difference between $\eta_R$ and $\eta_I$ is more than $100$ keV. 
In case of interaction with Higgs boson, which is elastic scattering, we can avoid the constraints if the Higgs coupling would be less than ${\cal O} (10^{-3})$. 

This paper is organized as follows. 
In Sec.~\ref{sec:II}, we review our model starting from the multiplication rules for our invertible symmetry, then constructing our valid Lagrangian, Higgs potential, radiative Dirac-type Yukawa terms, lepton flavor violations (LFVs), and muon $g-2$. 
In Sec.~\ref{sec:III}, we discuss our DM candidates; $N_{R_1}$, $S_0$, and $\eta_R$. 
Then, we demonstrate several numerical results for four cases of NH-$S_0$ DM, IH-$S_0$ DM, NH-$\eta_R$ DM, IH-$\eta_R$ DM in Sec.~\ref{sec:IV}, satisfying neutrino observables, LFVs, muon $g-2$, relic density of DM. 
Finally, we summarize and conclude in Sec.~\ref{sec:summary}.

\section{Model setup}
\label{sec:II}

Our model is to generate $y_\eta$ at one-loop level. 
In order to achieve it, we introduce three isospin doublet vector-like fermions $L'$ and singlet neutral boson $S_0$, as well as $N_R$ and $\eta$ in the Ma model. 
In addition, we impose an invertible symmetry in the class under ${\mathbb Z_2}$ gauging of ${\mathbb Z_6}$ symmetry, which is denoted by $\mathcal{FR} (6)$, to each field. 
The fusion rules are as follows:
\begin{align}
\begin{array}{lll}
\epsilon \otimes \epsilon = \mathbbm{1} \oplus \sigma \, , ~~~~~ & \sigma \otimes \sigma = \mathbbm{1} \oplus \sigma \, , ~~~~~ & \rho \otimes \rho = \mathbbm{1} \, , \\[0.3ex]
\epsilon \otimes \sigma = \epsilon \oplus \rho \, , ~~~~~ & \epsilon \otimes \rho = \sigma \, , ~~~~~ & \sigma \otimes \rho = \epsilon \, .
\end{array}
\end{align}
Therefore, following combinations of three fields
\begin{align}
\epsilon \otimes \epsilon \otimes \sigma \, , \quad \sigma \otimes \sigma \otimes \sigma \, , \quad \epsilon \otimes \sigma \otimes \rho \, ,
\label{eq:singletin3F}
\end{align}
and those of four fields
\begin{align}
\begin{array}{llll}
\epsilon \otimes \epsilon \otimes \epsilon \otimes \epsilon \, , ~~~~ & \epsilon \otimes \epsilon \otimes \epsilon \otimes \rho \, , ~~~~ & \epsilon \otimes \epsilon \otimes \sigma \otimes \sigma \, , ~~~~ & \epsilon \otimes \epsilon \otimes \rho \otimes \rho \, , \\[0.3ex]
\epsilon \otimes \sigma \otimes \sigma \otimes \rho \, , ~~~~ & \sigma \otimes \sigma \otimes \sigma \otimes \sigma \, , ~~~~ & \sigma \otimes \sigma \otimes \rho \otimes \rho \, , ~~~~ & \rho \otimes \rho \otimes \rho \otimes \rho \, ,
\end{array}
\label{eq:singletin4F}
\end{align}
are invariant under the $\mathcal{FR} (6)$. Note here that all elements are commutable each other. 
The former controls the Yukawa and scalar trilinear couplings, and the latter for the scalar quartic couplings. 

\begin{table}[!ht]
\begin{center}
\begin{tabular}{|c||c|c|c|c|c||c|c|c|}\hline\hline
& \multicolumn{5}{c||}{Fermions} & \multicolumn{3}{c|}{Scalars} \\ \hline
Fields & $L_L$ & $\ell_R$ & $N_R$ & $L'_L$ & $L'_R$ & $H$ & $\eta$ & $S_0$ \\ \hline
$SU(2)_L$ & $\mathbbm{2}$ & $\mathbbm{1}$ & $\mathbbm{1}$ & \multicolumn{2}{c||}{$\mathbbm{2}$} & $\mathbbm{2}$ & $\mathbbm{2}$ & $\mathbbm{1}$ \\ \hline
$U(1)_Y$ & $- \frac{1}{2}$ & $-1$ & $0$ & \multicolumn{2}{c||}{$- \frac{1}{2}$} & $\frac{1}{2}$ & $\frac{1}{2}$ & $0$ \\ \hline
$\mathcal{FR} (6)$ & $\mathbbm{1}$ & $\mathbbm{1}$ & $\rho$ & \multicolumn{2}{c||}{$\sigma$} & $\mathbbm{1}$ & $\epsilon$ & $\sigma$ \\ \hline
\end{tabular}
\caption{The charge assignments of relevant particles. 
Other particles not listed in this table have same SM charges with singlet under $\mathcal{FR} (6)$. }
\label{tab:chaged-assignments}
\end{center}
\end{table}
In Table~\ref{tab:chaged-assignments}, we summarize all relevant particle contents as well as its charge assignments. 
Note that the quarks are all singlet under $\mathcal{FR} (6)$. 
According to this charge assignments, we add the following terms to Eq.(\ref{eq:ma}):
\begin{align}
f_{ib} \overline{L_{L_i}} L'_{R_b} S_0 + f'_{ab} \overline{L'_{L_a}} L'_{R_b} S_0 + g_{ab} \overline{L'_{L_a}} N_{R_b} \widetilde{\eta} + h_{ab} \overline{L^{\prime c}_{R_a}} N_{R_b} \eta 
+ M_{L'_b} \overline{L'_{L_b}} L'_{R_b} + {\rm h.c.} , \label{eq:ma_1}
\end{align}
where $M_{L'}$ is diagonal without loss of generality. 
The scalar potential is found as follows:
\begin{align}
V &= \sum_{\phi = H, \eta, S_0} \Bigl[ - \mu_{\phi}^2 |\phi|^2 + \lambda_{\phi} |\phi|^4 \Bigr] + \kappa_S S_0^3 + \left( \mu |\eta|^2 S_0 + \lambda''_{H \eta} (H^{\dagger} \eta)^2 + {\rm c.c.} \right) \nonumber \\[0.3ex]
&\hspace{1.2em} + \lambda_{H \eta} |H|^2 |\eta|^2 + \lambda'_{H \eta} |H^{\dagger} \eta|^2 + \lambda_{H S} |H|^2 |S|^2 + \lambda_{\eta S} |\eta|^2 |S|^2 \, .
\label{eq:ma_2}
\end{align}
Below, we move on to how to generate the Yukawa coupling $y^\eta$ instead of discussing the scalar potential.

\subsection{$y^\eta$ at one-loop level}
\label{sec:yeta1loop}

$y^\eta$ is induced via terms $f$ and $g$ in Eq.~(\ref{eq:ma_1}) and $\mu$ in Eq.~(\ref{eq:ma_2}) at one-loop level, which is calculated as follows:
\begin{align}
y^\eta_{ia} &=\frac{\mu}{(4\pi)^2(m_S^2 - m_0^2)} 
\sum_{b=1}^3 {f_{ib} M_{L'_b} g_{ba}}
\left(
\frac{m_S^2}{m_S^2-M_{L'_b}^2} \ln\left[\frac{m_S^2}{M_{L'_a}^2}\right]
-
\frac{m_0^2}{m_0^2-M_{L'_a}^2} \ln\left[\frac{m_0^2}{M_{L'_a}^2}\right]
\right)\\
&\equiv f_{ib} \widetilde{D'_{b}} g_{ba}.
\label{eq:y_eta-3lp}
\end{align}
where $m_0\equiv (m_R+m_I)/2$. 
From Eq.~(\ref{eq:y_eta}) and Eq.~(\ref{eq:y_eta-3lp}), we can rewrite $f$ and $g$ as
\begin{align}
f= U^*\sqrt{ D_\nu} \left({\widetilde{D'}}\right)^{-1/2},\\
g= \left({\widetilde{D'}}\right)^{-1/2} O_N D_N^{-1/2},
\end{align}
where Max$[|f|]\lesssim4\pi$ and Max$[|g|]\lesssim4\pi$.

\subsection{LVFs and muon $g-2$}
\label{sec:leptonpheno}

New contributions for LFVs arise from $f$ and their formulae are simply given by changing $y^\eta\to f,\ m_\eta\to m_S,\ M_N\to M_{L'}$ in Eqs.~\eqref{eq:aR} and \eqref{eq:aL} and flipping the overall sign~\footnote{Sign flip comes from a fact that the photon attaches charged-fermions instead of charged-bosons.}. 
Therefore, 
\begin{align}
&(a_R)_{ij} \approx - \frac{f_{ja} f^\dag_{ai} m_{\ell_i}}{12(4\pi)^2} 
\left[\frac{M_{L'_a}^6 - 6 M_{L'_a}^4 m^2_S + 3M_{L'_a}^2 m^4_S +2 m^2_S+6 M_{L'_a}^2 m^4_S\ln\left(\frac{M_{L'_a}^2}{m^2_S}\right)}
{(M_{L'_a}^2 - m^2_S)^4}\right], \label{eq:aR_3lp}\\
&(a_L)_{ij} \approx - \frac{f_{ja} f^\dag_{ai} m_{\ell_j}}{12(4\pi)^2} 
\left[\frac{M_{L'_a}^6 - 6 M_{L'_a}^4 m^2_S + 3M_{L'_a}^2 m^4_S +2 m^2_S+6 M_{L'_a}^2 m^4_S\ln\left(\frac{M_{L'_a}^2}{m^2_S}\right)}
{(M_{L'_a} - m^2_S)^4}\right].\label{eq:aL_3lp}
\end{align}
From Eq.~\eqref{eq:mug-2}, this sign flipping leads to positive muon $g-2$, which is clear difference from the Ma model.

\section{Dark matter candidates}
\label{sec:III}

In our model, we have extra DM candidate from the Ma model, which is $S_0$. 
Before showing our numerical results, we briefly summarize each feature in this section.

\subsection{$N_{R_1}$ DM}
\label{sec:NRDM}

If we consider the fermion DM candidate, only the source to explain the relic density of DM is obtained via Yukawa couplings $y^\eta$, as mentioned in the introduction. 
However since $y^\eta$ is generated at one-loop level, that would be tiny. 
In fact, we have found the maximum component of $y^\eta$ is at most $10^{-4}$ for both the cases of NH and IH through our numerical analysis. 
On the other hand, the observed relic density of DM requires order one of $y^\eta$. 
Thus, the lightest mass of the neutral fermion cannot be a dominant component of the DM, and hence, we omit the numerical analysis for the fermion DM case in this paper.

\subsection{$S_0$ DM}
\label{sec:S0DM}

Here, we consider the isospin singlet DM candidate $S_0$. 
For this purpose, we impose the following mass hierarchy
\[
m_S < m_R,m_I,M_{N_1},
\]
where $m_R, m_I, M_{N_1}$ are masses of the other DM candidates and their analyses are done by several papers~\cite{Kubo:2006yx}. 
At first, we suppose the main annihilation cross section to explain the relic density of DM is arisen from Yukawa interaction via $f$. 
Moreover, the cross section can be expanded by relative velocity of $v_{\rm rel}$:
$\sigma v_{\rm rel}\approx a_{\rm eff}+ b_{\rm eff} v_{\rm rel}^2+ d_{\rm eff} v_{\rm rel}^4+{\cal O}(v_{\rm rel}^6)$. 
Each coefficient is given by
\begin{align}
& a_{\rm eff} =\sum_{b=1}^3\frac{f_{3b}f^\dag_{b3}}{8\pi}\frac{m_\tau^2}{(m_S^2+M_{L'_b}^2-m^2_\tau)^2} \theta(m_S-m_\tau), \\
& b_{\rm eff} =\sum_{b=1}^3\frac{f_{3b}f^\dag_{b3}}{24\pi}\frac{m_S^2 m_\tau^2(m_\tau^2-3 M^2_{L'_b}-m_S^2)}{(m_S^2+M_{L'_b}^2-m^2_\tau)^4} \theta(m_S-m_\tau),\\
& d_{\rm eff} \approx\sum_{i,j=1}^3
\sum_{b=1}^3\frac{f_{ib}f^\dag_{bj}+{\hat{f}_{ib}}{\hat{f}^\dag_{bj}}}{120\pi}\frac{m_S^6 }{(m_S^2+M_{L'_b}^2)^4},
\end{align}
where $\hat{f}\equiv U^\dag f$. 
The DM is d-wave dominant in the limit of zero mass of final states. 
The typical range of the cross section is given by
\begin{align}
1.70630\le (\sigma v_{rel}) \times 10^9\ {\rm GeV}^2 \le 2.03102,
\end{align}
where it corresponds to $0.120 \pm 2 \times 0.001$~\cite{Planck:2018vyg}. 
However, we have found that the DM would not reach the above cross section through our numerical analysis~\footnote{The annihilation is ${\cal O}(10^{-28})$ GeV$^{-2}$ at most that is far below the required cross section.}. 
Thus, we need to rely on the other interactions arising from Higgs potential and kinetic terms. 
{\it One notice here is that $S_0$ can decay into a pair of the SM Higges via $\mu$ and $\lambda_{H\eta}$ at one-loop level. 
Thus, we need to make them tiny enough to evade the too-short life time of $S_0$ when $250 \, {\rm GeV} \lesssim m_S$. }

\subsection{$\eta_R$ DM}
\label{sec:etaDM}

In case of $\eta_R$ DM, the main cross section would come from kinetic terms~\cite{Hambye:2009pw}, since $y^\eta$ is too small to explain the relic density as discussed in $N_{R_1}$ DM candidate. 
Thus, the situation is the same as the case of original Ma model and we can adopt the same condition as can be seen in the introduction:
\begin{align}
m_R\approx 534 \pm 8.5 \ {\rm GeV}\ ({\rm 1\sigma}).
\end{align}
$\eta$ interacts with quark sector via $Z$-boson or Higgs boson at tree-level, and this case will suffer from the constraints of direct detections. 
As mentioned in the introduction, however, the constraints can be easily evaded by the enough mass difference between $\eta_R$ and $\eta_I$ ($\ge 100$ keV) for the $Z$-boson interaction case and small Higgs coupling ($\le {\cal O} (10^{-3})$) for the Higgs boson interaction case. 
{\it Unlikely to the $S_0$ DM case, $\eta$ never decay into the SM particles. This is because it is assured by ${\mathbb Z_2}$ symmetry originated from $\mathcal{FR} (6)$. }

\section{Numerical results}
\label{sec:IV}

In this section, we perform numerical analyses to satisfy the lepton masses and mixing angles, LFVs and muon $g-2$ where we adopt the best fit values of neutrino oscillation data in Nufit 6.0~\cite{Esteban:2024eli} without Super-Kamiokande atmospheric data and charged-lepton masses in PDG~\cite{ParticleDataGroup:2022pth}. 
Then, we check the allowed parameter space can also satisfy the relic density of DM for the cases of $S_0$ and $\eta_R$ below. 
Before discussing the numerical analysis, it would be worthwhile mentioning the results of neutrino mass observables: $\sum D_\nu$, $m_{ee}$, $m_{\nu e}$, discussed in the introduction. 
When we impose $\sum D_\nu\le120$ meV and randomly select the lightest neutrino mass eigenvalue in the range of $D_{\nu1}(D_{\nu3})\le1$ eV and two Majorana phases in the range of ($-\pi,\pi$), we find the following upper bounds:
\begin{align}
{\rm NH}&:\ \{D_{\nu_1}, m_{ee}, m_{\nu e}\} \lesssim \{30.031, \ 30.451, \ 31.305\} \ {\rm meV}, \\
{\rm IH}&:\ \{D_{\nu_3}, m_{ee}, m_{\nu e}\} \lesssim \{15.707, \ 50.910, \ 51.480\} \ {\rm meV}.
\end{align}

\subsection{Numerical results of $S_0$ DM}
\label{sec:III_s0}

We consider the DM candidate in case of $S_0$ and randomly select our input parameters in the following ranges:
\begin{align}
&\{m_S, \mu\} = [0.1-10^3] \, {\rm GeV}, \ m_\eta = [m_S-10^3] \, {\rm GeV}, \ \{M_{N_1} \le M_{N_2} \le M_{N_3}\} = [m_S-10^3] \, {\rm GeV},\\
&\{M_{L'_1} \le M_{L'_2} \le M_{L'_3}\} = [m_S-10^3] \, {\rm GeV}, \ |\theta_{12, 23, 13}| = [0.0-\pi],
\end{align}
where $\theta_{12, 23, 13}$ are mixing angles of $O_N$. 

\subsubsection{\rm NH}
 
\begin{figure}[tb]
\begin{center}
\includegraphics[width=53mm]{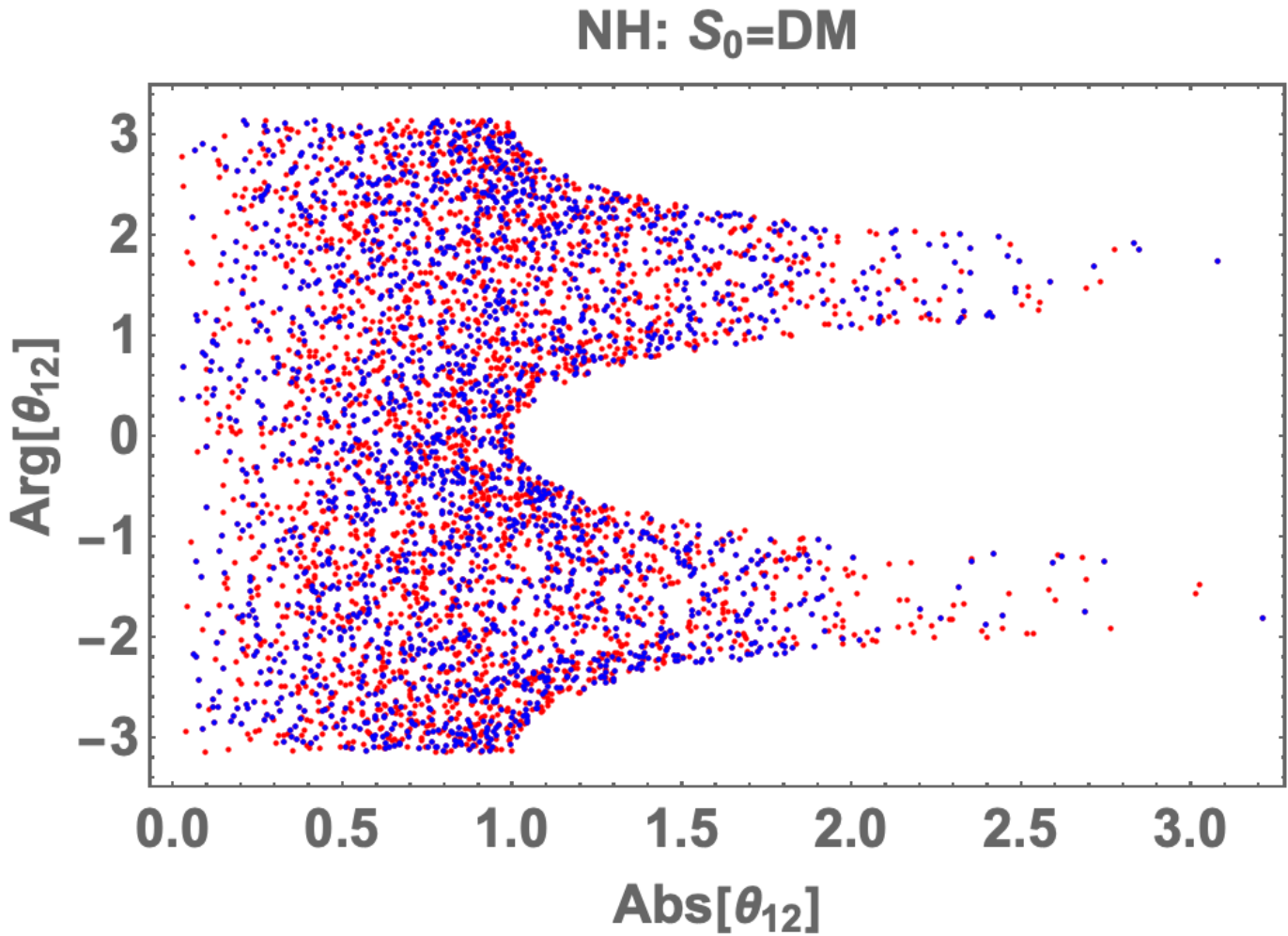}
\includegraphics[width=53mm]{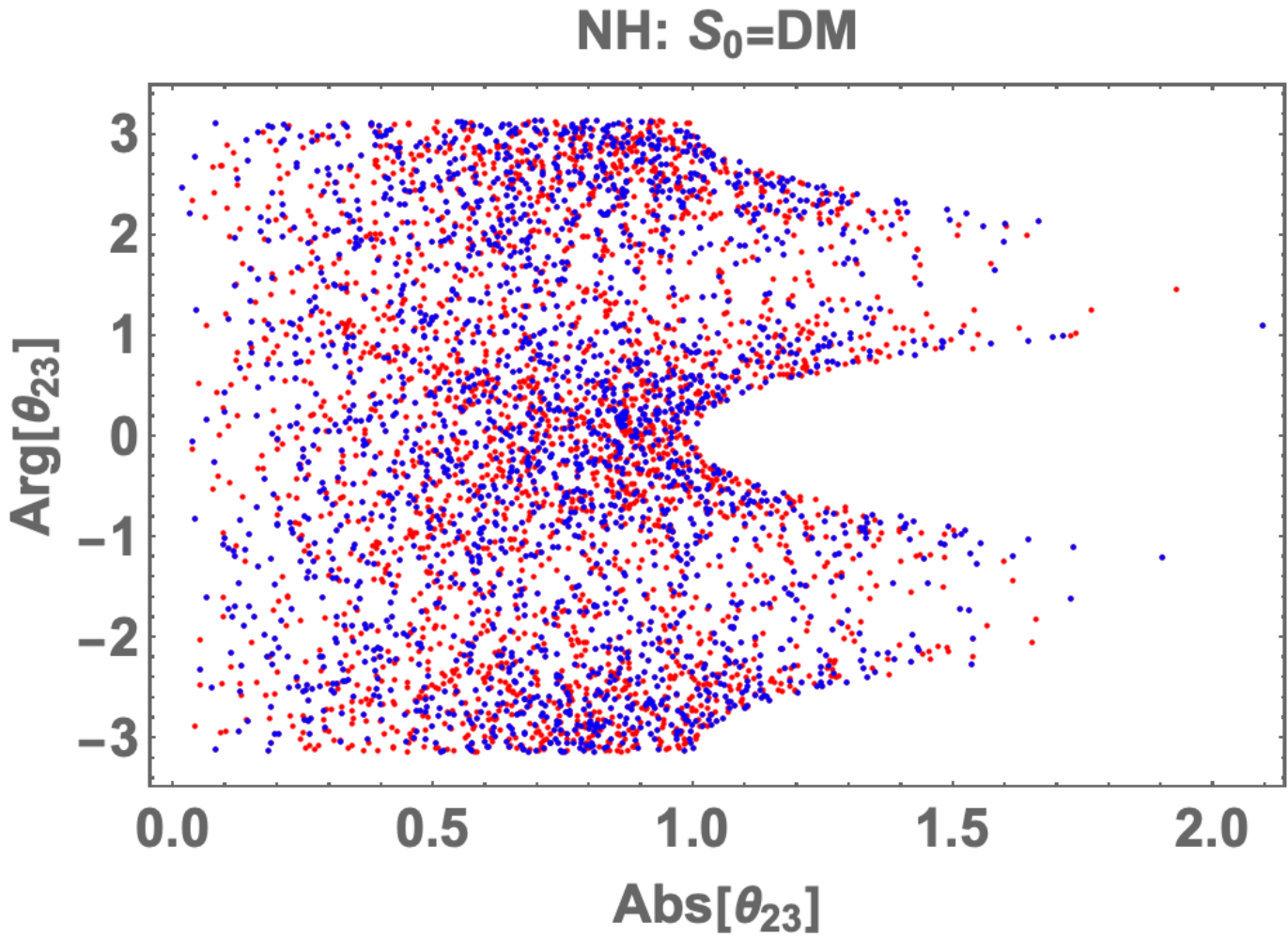}
\includegraphics[width=53mm]{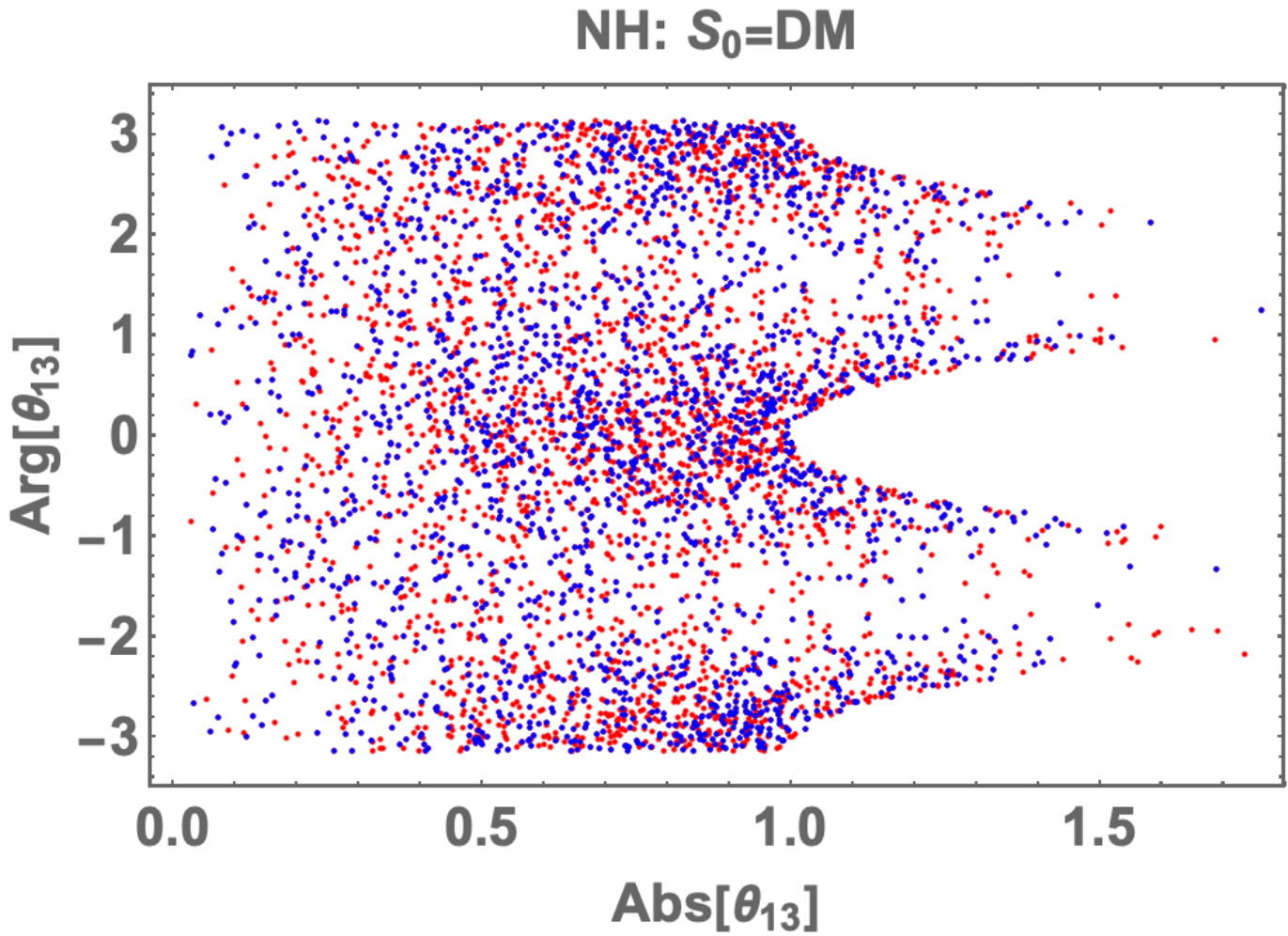}
\caption{Allowed regions for $\{\theta_{12}, \theta_{23}, \theta_{13}\}$, for the NH case with $S_0$ DM. 
Blue points are allowed only when $\sum D_\nu \le 120$ meV is satisfied, otherwise points are colored by red. 
The vertical axis represents arguments, while the horizontal axis does the absolute values for these mixing angles. } 
\label{fig:nh1_s}
\end{center}
\end{figure}
\begin{figure}[tb]
\begin{center}
\includegraphics[width=88mm]{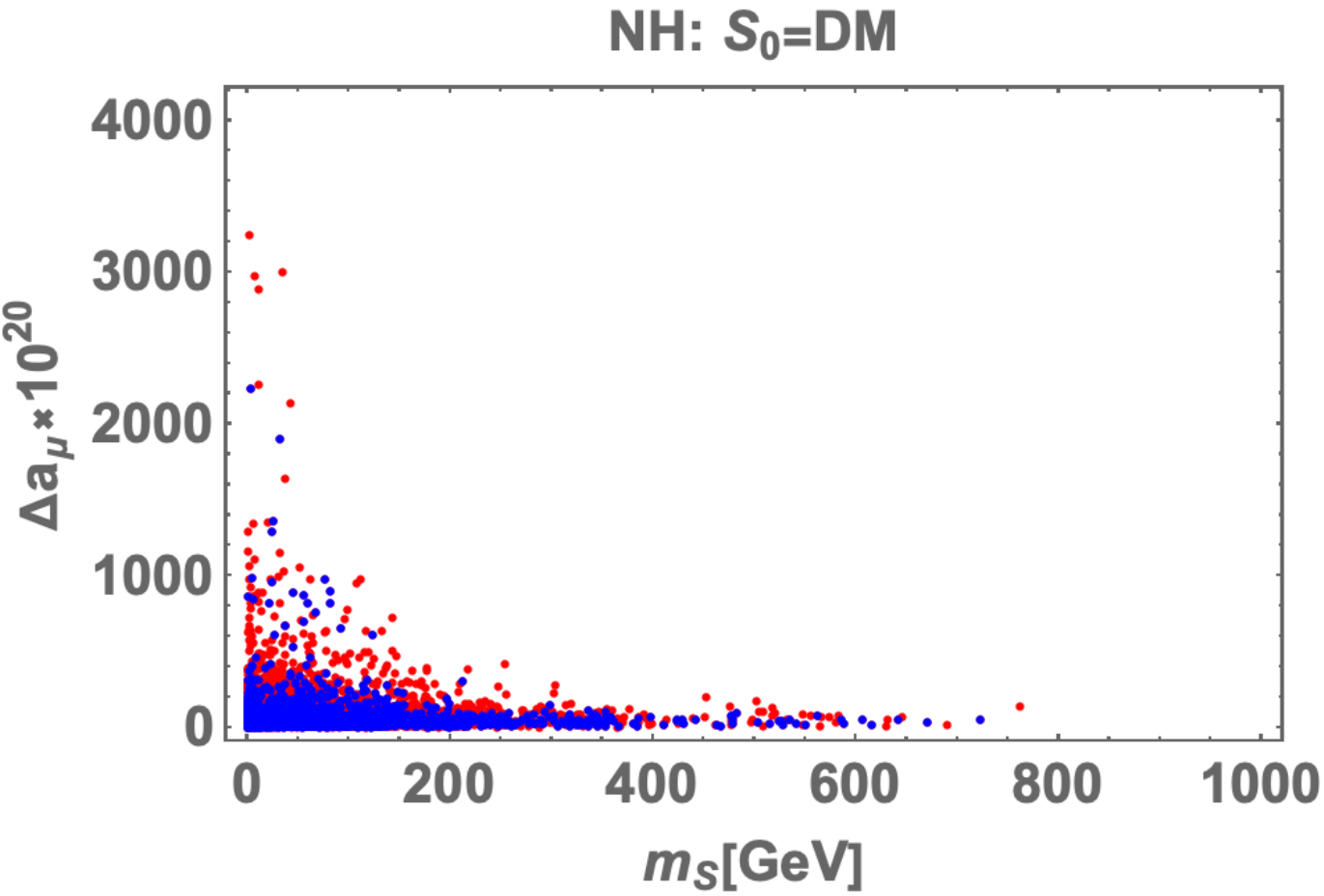}
\caption{Allowed regions for the muon $g-2$ in terms of the DM mass, for the NH case with $S_0$ DM. 
All color legends are the same as the ones in Fig.~\ref{fig:nh1_s}. } 
\label{fig:nh2_s}
\end{center}
\end{figure}
Fig.~\ref{fig:nh1_s} shows allowed regions for $\theta_{12}$ (left), $\theta_{23}$ (center), $\theta_{13}$ (right) for the NH case. 
Blue points are allowed only when $\sum D_\nu \le 120$ meV is satisfied, otherwise points are colored by red. 
Hereafter, we use this color legend for similar plots, without mentioning. 
The vertical axis represents arguments, while the horizontal axis does the absolute values for these mixing angles. 
These figures imply that they are unique shapes: all ranges for arguments are allowed for Abs$[\theta_{ij}] < 1.0$, while only specific regions are allowed for Abs$[\theta_{ij}] > 1.0$. 

Fig.~\ref{fig:nh2_s} shows allowed regions for muon $g-2$ in terms of the mass of DM. 
As can be seen the figure, the predicted muon $g-2$, $\Delta a_\mu \lesssim 3.2 \times 10^{-17}$, has almost no discrepancy from the SM prediction, and therefore, the result is consistent with the current discrepancy between theoretical and experimental results in Eq.~\eqref{eq:Delamu}. 
This figure shows another feature for our model: the upper bound on the DM mass is about 800 GeV. 

\subsubsection{\rm IH}

\begin{figure}[tb]
\begin{center}
\includegraphics[width=53mm]{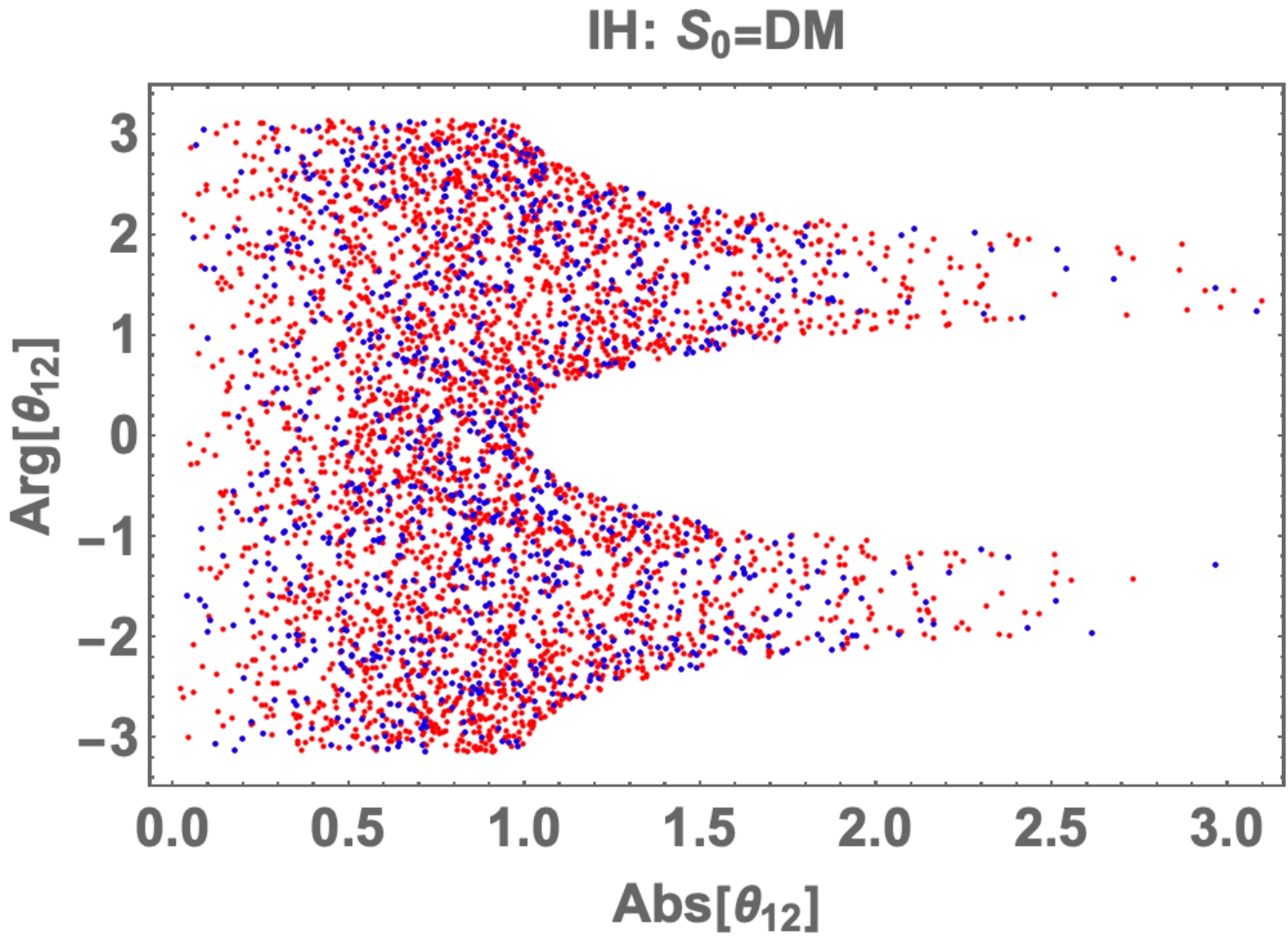}
\includegraphics[width=53mm]{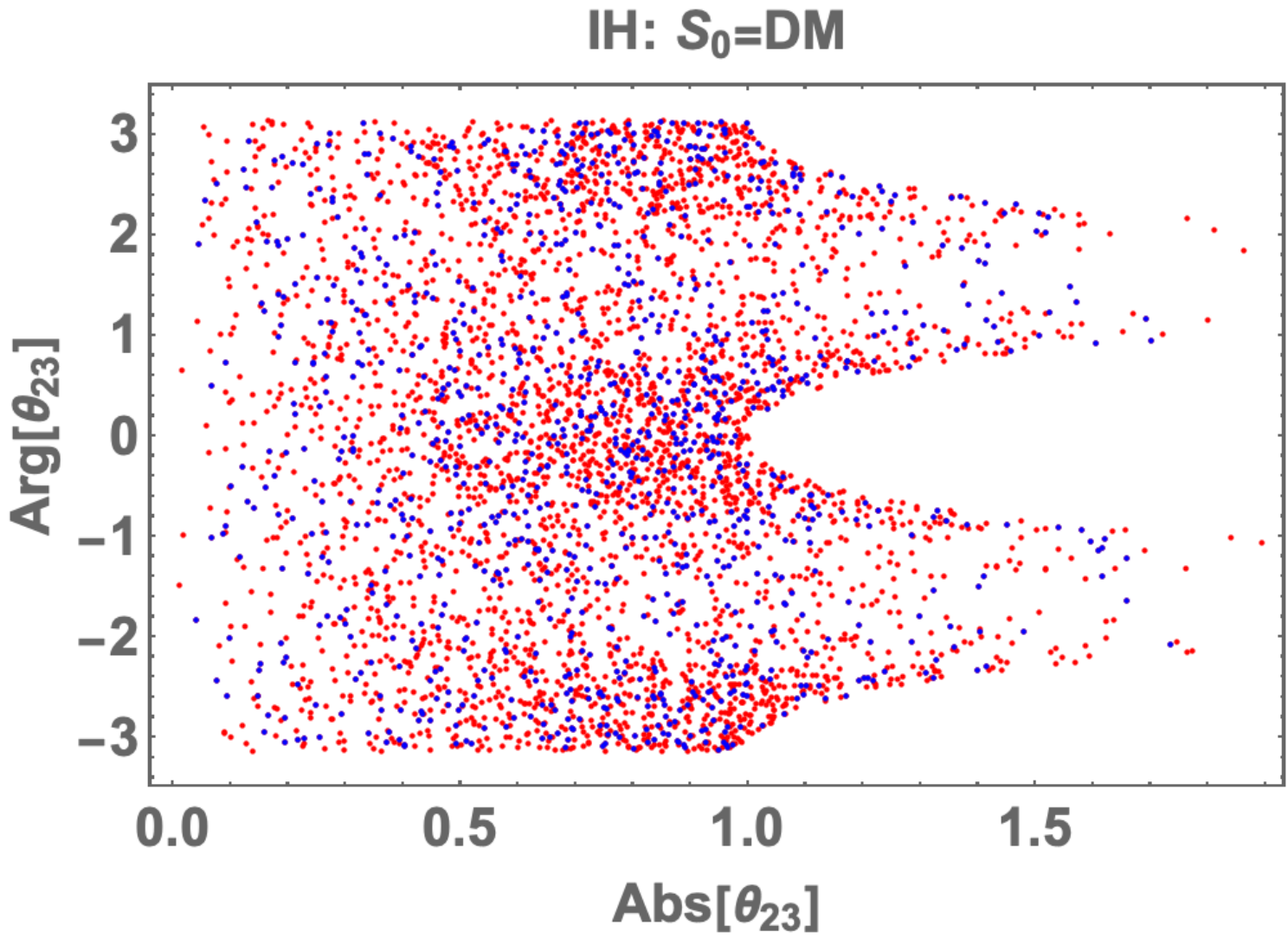}
\includegraphics[width=53mm]{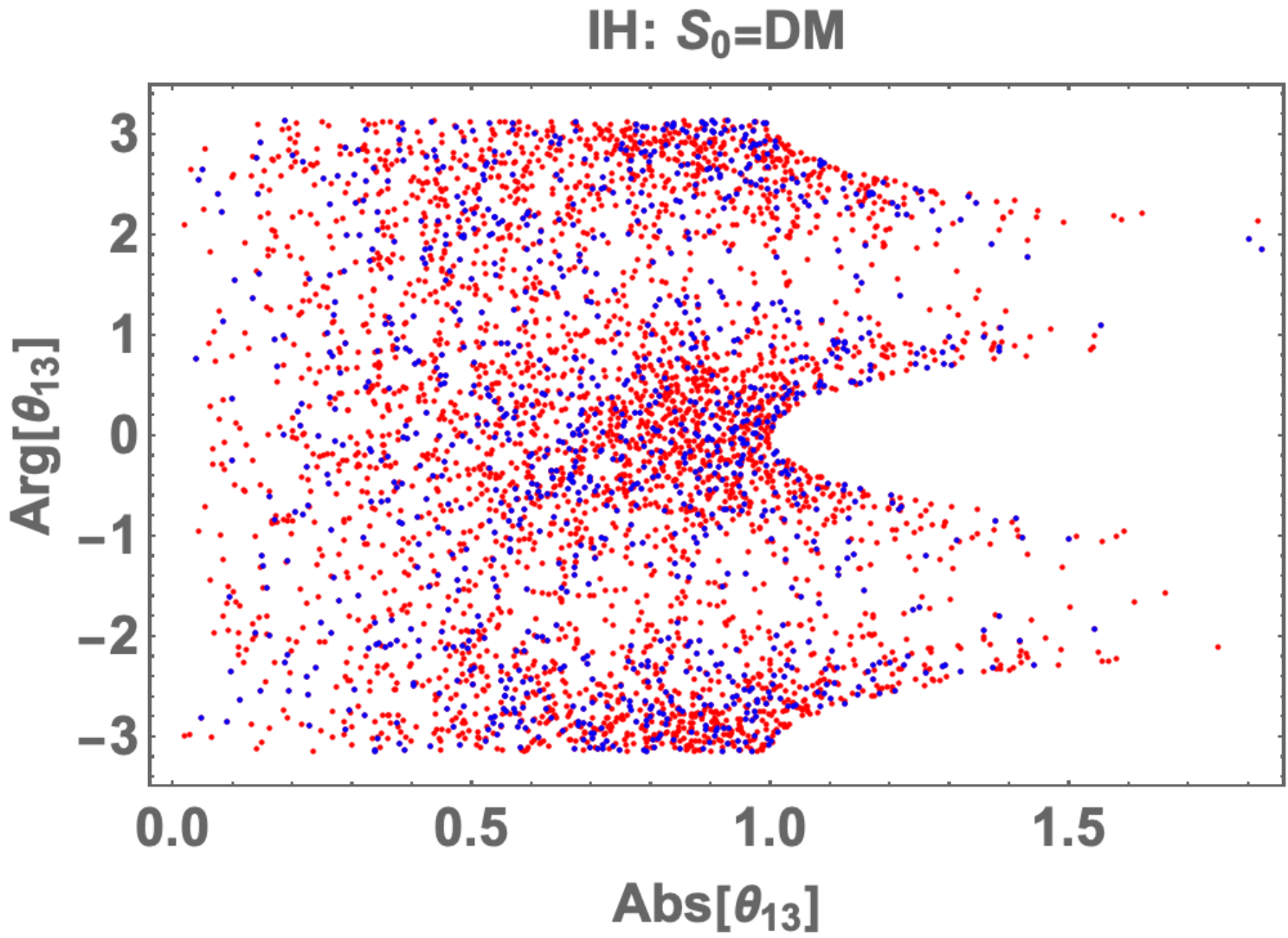}
\caption{Same plots as Fig.~\ref{fig:nh1_s}, for the IH case with $S_0$ DM. }
\label{fig:ih1_s}
\end{center}
\end{figure}
\begin{figure}[tb]
\begin{center}
\includegraphics[width=88mm]{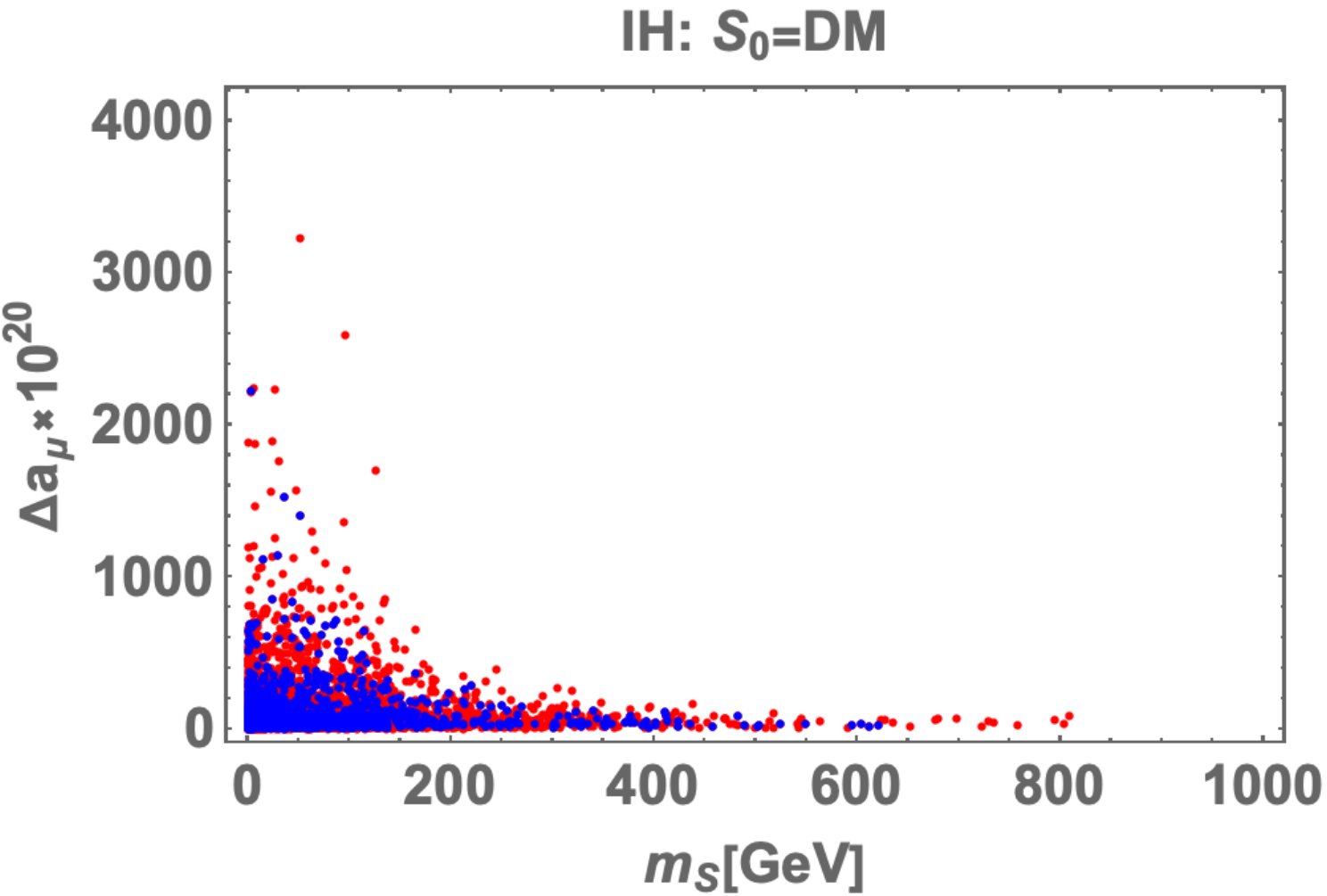}
\caption{Same plot as Fig.~\ref{fig:nh2_s}, for the IH case with $S_0$ DM. } 
\label{fig:ih2_s}
\end{center}
\end{figure}
Fig.~\ref{fig:ih1_s} shows allowed regions for $\theta_{12}$ (left), $\theta_{23}$ (center), $\theta_{13}$ (right) for the IH case. 
Comparing with Fig.~\ref{fig:nh1_s}, the shapes of allowed regions are similar with each other, while more points are out of $\sum D_\nu \le 120$ meV. 

Fig.~\ref{fig:ih2_s} shows allowed regions for the muon $g-2$ in terms of the mass of DM. 
Similar to the NH case, our muon $g-2$ has almost no discrepancy from the SM prediction: $\Delta a_\mu \lesssim 3.2 \times 10^{-17}$, and the similar upper bound on the DM mass is obtained, $m_S \lesssim 800$ GeV. 

\subsubsection{DM phenomenology}

For the calculation of the relic density from $S_0$ DM, we simply neglect all Yukawa couplings in the calculation, and hence, the following results are common both for the NH and IH cases, as long as the contributions from these Yukawa couplings are enough small. 
We choose the relevant parameters as
\begin{align}
&m_{\eta} = [m_S - 10^3] \, {\rm GeV} \, , \quad m_S = [0 - 800] \, {\rm GeV} \, , \quad \{\kappa_S, \mu\} = [0 - 10^3] \, {\rm GeV} \, , \\[0.3ex]
&\lambda_{H \eta} = 0.1 \, , \quad \lambda_{H S} = 0.001 \, , \quad \lambda_{\eta S} = 0.0001 \, ,
\end{align}
and generate about $10^5$ samples randomly. 
Here, the mass range for $m_S$ is set be consistent with those obtained in previous results, and $\kappa_S$ is a parameter for $S_0^3$ coupling, defined in Eq.~\eqref{eq:ma_2}. 
Note that other mass parameters $m_{\eta^{\pm}}, m_I$ are set to be equal to $m_{\eta} = m_R$ for simplicity, which corresponds to the case of $\lambda'_{H \eta} = \lambda''_{H \eta} = 0$. 
Remaining quartic couplings $\lambda_{\eta}, \lambda_S$ are totally irrelevant for the relic density calculation. 

\begin{figure}[tb]
\begin{center}
\includegraphics[width=80mm]{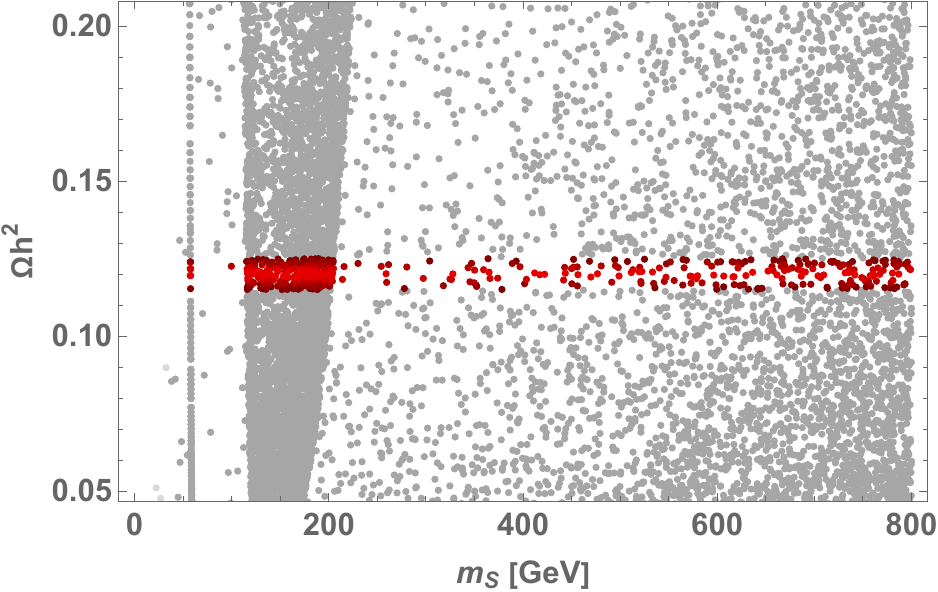} ~~
\includegraphics[width=80mm]{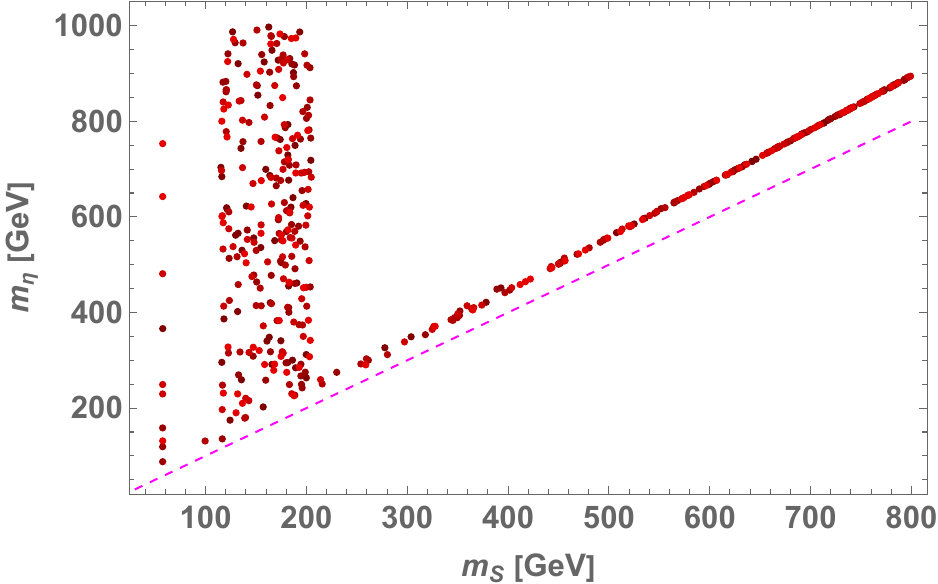}
\caption{Plots for the relic density in the case of the $S_0$ DM. 
\textit{Left}: predicted relic density for each value of $m_S$; \textit{right}: relic density in the $(m_S, m_{\eta})$-plane. 
We plot the DM relic density of $\Omega h^2 = 0.120 \pm 0.001$~\cite{Planck:2018vyg} within $1\sigma, ..., 5 \sigma$ (in gradations of red from lighter to darker) and outside of $5\sigma$ (gray). 
In the right panel, we only plot the relic density within $5\sigma$, and the magenta dashed line shows $m_S = m_{\eta}$. 
In both panels, the direct detection constraints are satisfied all points. }
\label{fig:DM_s}
\end{center}
\end{figure}
Fig.~\ref{fig:DM_s} shows our numerical results by using \texttt{micrOMEGAs}~\cite{Belanger:2018ccd,Belanger:2020gnr,Belanger:2021smw,Alguero:2022inz,Alguero:2023zol}~\footnote{For more details, see \href{https://lapth.cnrs.fr/micromegas/}{https://lapth.cnrs.fr/micromegas/}.}, in order for calculating the relic density as well as checking the direct detection bounds. 
In the left panel, the values of $\Omega h^2$ for each DM mass are shown. 
The gray dots correspond to the case where predicted $\Omega h^2$ is outside of $5\sigma$ of current result, $\Omega h^2 = 0.120 \pm 0.001$, while red plots indicate the prediction within $1\sigma, ..., 5\sigma$ in gradation from lighter to darker. 
The right panel shows the predictions in $(m_S, m_{\eta})$-plane, but only plot $\Omega h^2$ within $5\sigma$ cases. 
Here, the effect from the loop decay of $S_0$ mentioned above is not included in the calculation, and therefore, some points in the range of $250 \, {\rm GeV} \leq m_S$ may disappear due to the short lifetime. 
Note that in these plots, all points satisfy the direct detection bounds. 

From the right panel of Fig.~\ref{fig:DM_s}, there are several critical DM mass ranges for this calculation: (i) $m_S \lesssim 100$ GeV, (ii) $100 \, {\rm GeV} \lesssim m_S \lesssim 210$ GeV and (iii) $210 \, {\rm GeV} \lesssim m_S$. 
\begin{itemize}
\item[(i)] For this region, only $m_S \simeq 57.5$ GeV gives correct relic density, and the main annihilation mode is $S_0 \, S_0 \to b \, \bar{b}$ ($\simeq 86.5\%$), through the Higgs exchange. 
Sub-dominant ones are $S_0 \, S_0 \to \tau \, \bar{\tau}$ ($\simeq 9.1\%$) and $S_0 \, S_0 \to c \, \bar{c}$ ($\simeq 4.3\%$), which fixed by sizes of each SM Yukawa coupling. 

\item[(ii)] For this region, the main annihilation becomes the ``semi-annihilation"~\cite{DEramo:2010keq} of $S_0 \, S_0 \to S_0 \, h^0$ with the SM Higgs $h^0$, depending on the size of $m_{\eta}$. 
When $m_{\eta} - m_S > 50$ GeV, this semi-annihilation process is $\simeq 99.9\%$ of all relevant annihilation processes and determines the relic density of $S_0$. 
On the other hand, once the mass relation becomes $m_{\eta} - m_S < 50$ GeV, self-annihilation of $S_0 \, S_0 \to \eta^+ \eta^-$ and coannihilation of $S_0 \, \eta_{R, I} \to \eta^{\pm} W^{\mp}$ become relevant processes to the correct relic density. 
Note that for this region, how dominant process is obtained also depends on the size of $\kappa_S$ and $\mu$: enough size of $\kappa_S > \mu$ leads the self-annihilation to be dominant, while that of $\kappa_S < \mu$ leads the coannihilation dominant. 

\item[(iii)] For this region, it is clear that the correct relic density is realized by the self-annihilation processes $S_0 \, S_0 \to \eta^+ \eta^-$, or the coannihilation ones $S_0 \, \eta_{R, I} \to \eta^{\pm} W^{\mp}$, depending on the size of $\kappa_S$ and $\mu$. 
\end{itemize}
It is notable that these crucial DM mass ranges will be changed by choosing different choice for quartic couplings, $(\lambda_{H \eta}, \lambda_{H S}, \lambda_{\eta S})$. 
Among these, the size of $\lambda_{H S}$ should be enough small, otherwise the direct detection bounds exclude almost all the viable parameter space.

\subsection{Numerical results of $\eta$ DM }
\label{sec:eta}

Here, we consider the case of real part of $\eta$ as the DM candidate. 
We work on the following DM mass range suggested by the ref.~\cite{Hambye:2009pw}:
\begin{align}
m_R \approx 534 \pm 2 \times 8.5 \, {\rm GeV},
\end{align}
where $m_R$ is the mass of $\eta_R$. 
Then, we randomly select our input parameters in the following ranges:
\begin{align}
&\{m_S, \mu\} = [0-10^4] \, {\rm GeV}, \ \{M_{N_1} \le M_{N_2} \le M_{N_3}\} = [m_R-10^4] \, {\rm GeV}, \\
&\{M_{L'_1} \le M_{L'_2} \le M_{L'_3}\} = [m_R-10^4] \, {\rm GeV}, \ |\theta_{12, 23, 13}| = [0.0-\pi].
\end{align}

\subsubsection{\rm NH}

\begin{figure}[tb]
\begin{center}
\includegraphics[width=53mm]{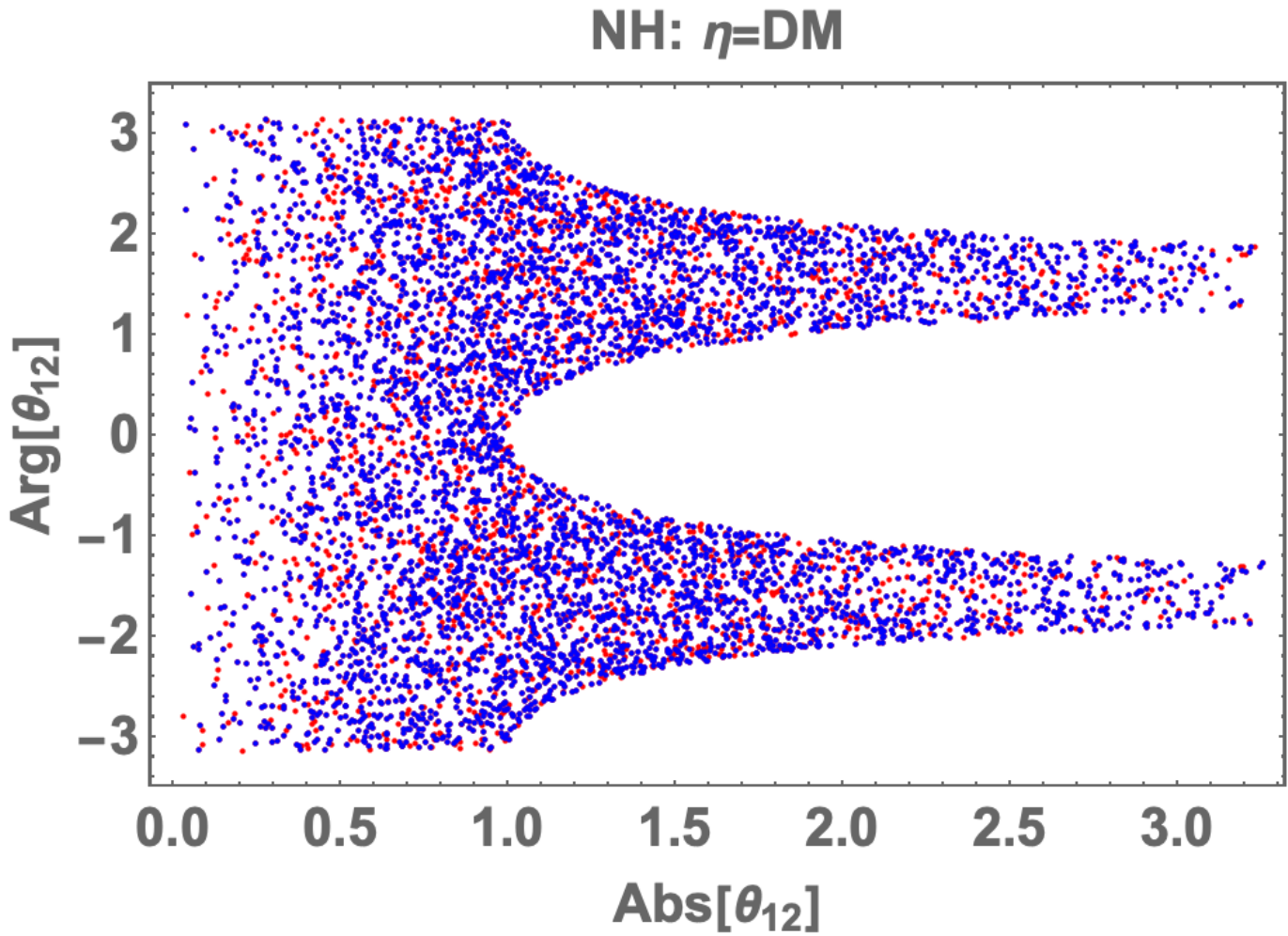}
\includegraphics[width=53mm]{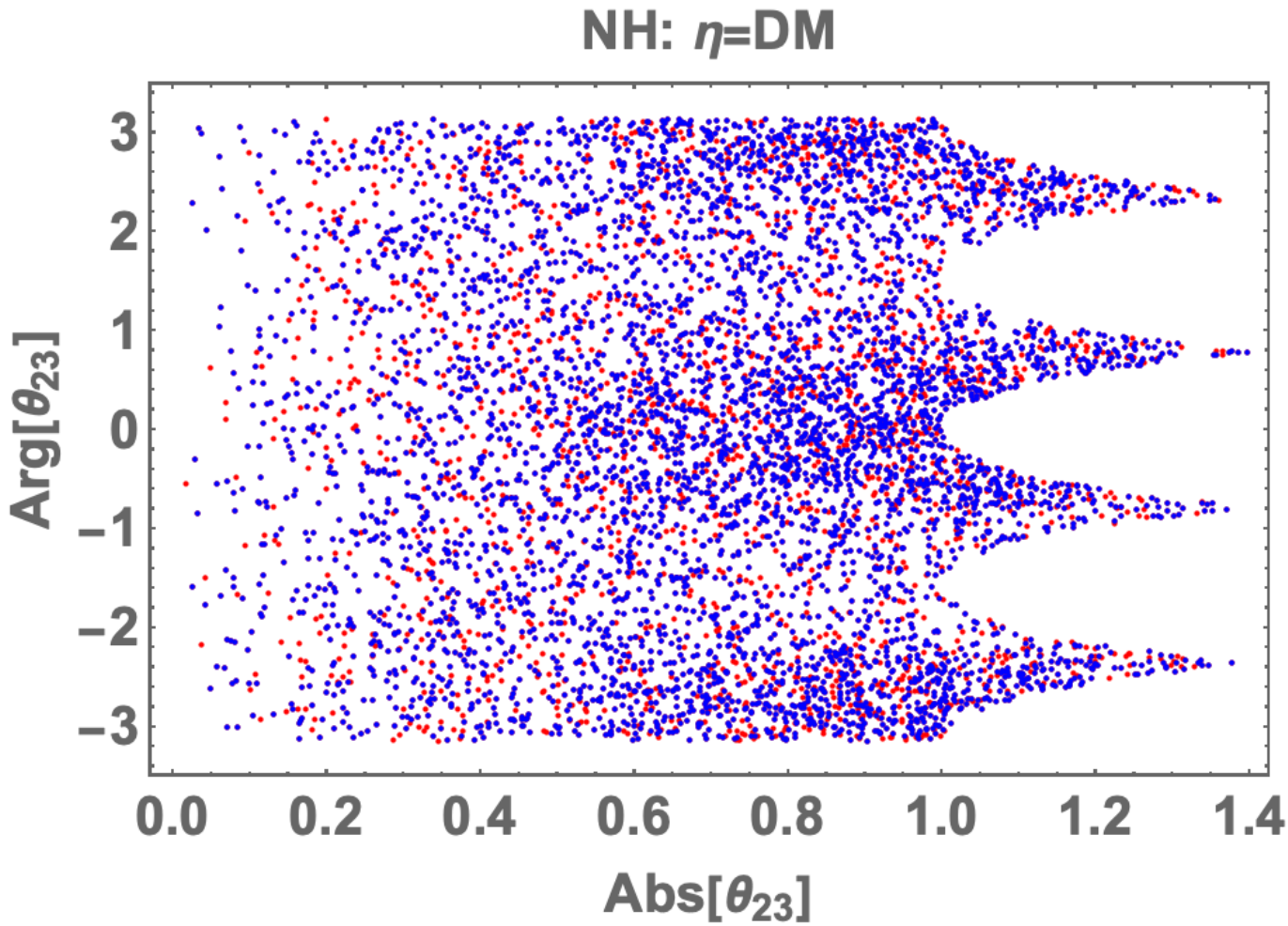}
\includegraphics[width=53mm]{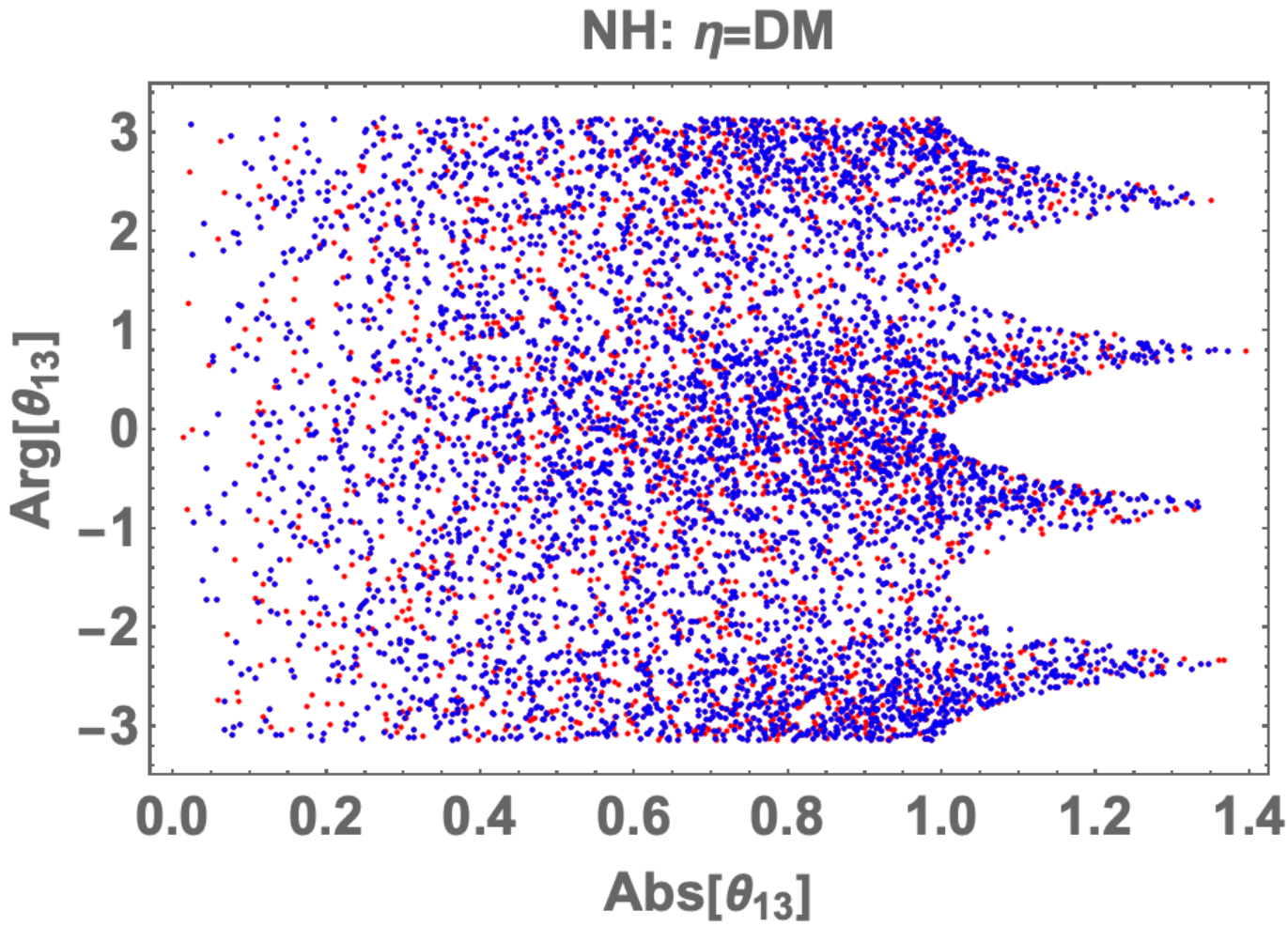}
\caption{Allowed regions for $\{\theta_{12}, \theta_{23}, \theta_{13}\}$, for the NH case with $\eta$ DM. }
\label{fig:nh1_eta}
\end{center}
\end{figure}
\begin{figure}[tb]
\begin{center}
\includegraphics[width=88mm]{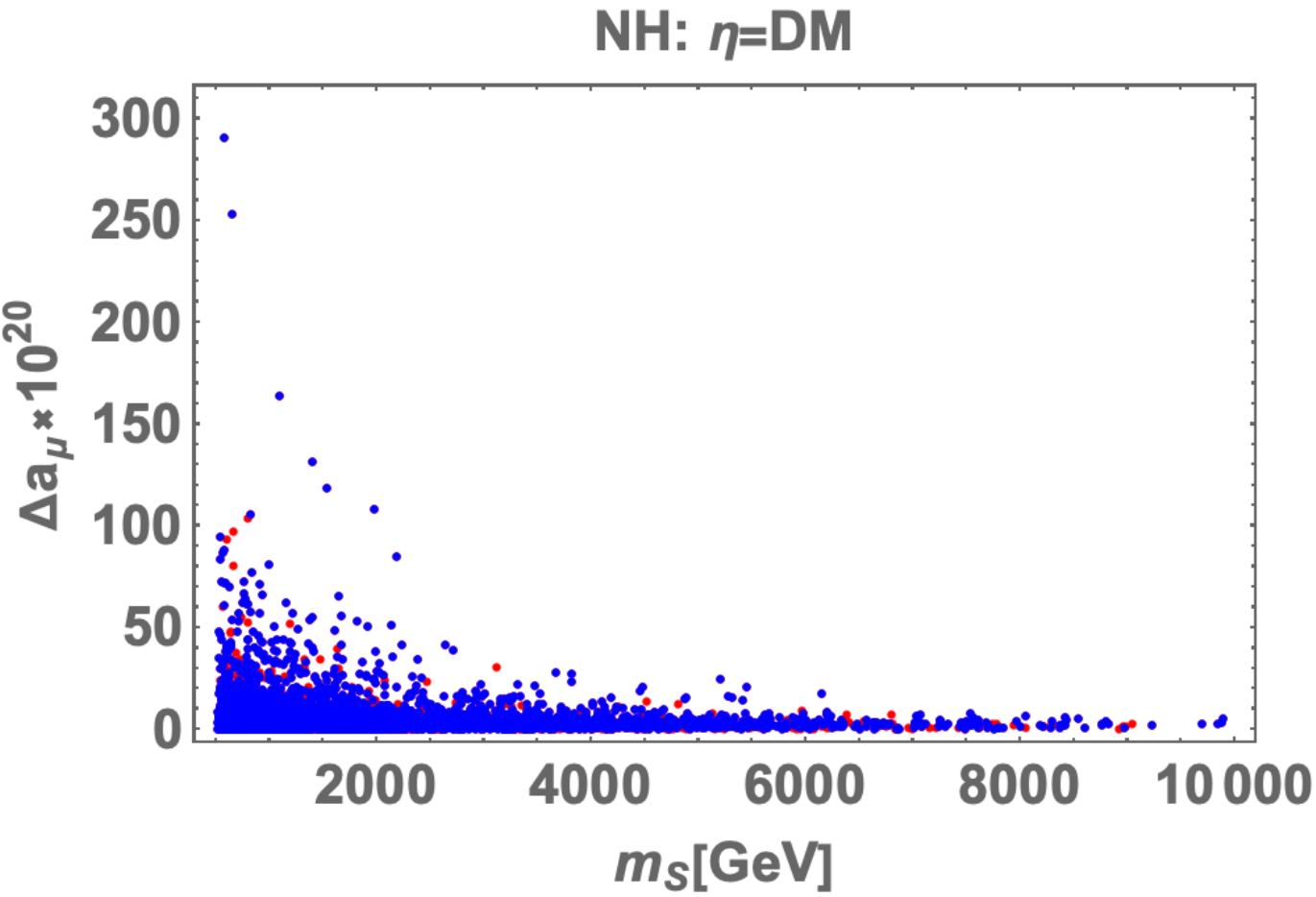}
\caption{Allowed regions for muon $g-2$ in terms of the mass of DM where all the color legends are the same as the one of Fig.~\ref{fig:nh1_s}. }
\label{fig:nh2_eta}
\end{center}
\end{figure}
Fig.~\ref{fig:nh1_eta} shows allowed regions for $\theta_{12}$ (left), $\theta_{23}$ (center), $\theta_{13}$ (right). 
The vertical axis represents arguments, while the horizontal axis does the absolute values for these mixing angles. 
These figures imply that allowed regions are similar to the $S_0$ DM case but there are two times of cusps for $\theta_{23,13}$ than the $S_0$ DM case. 
In addition, the maximum values of $\theta_{23, 13}$ are smaller than those for the $S_0$ DM case. 
Furthermore, we obtained many more blue points, compared with the $S_0$ DM case, both of the NH and IH cases. 
This will be originated from the heavy mass of $m_S$, which is $m_S \gtrsim 534$ GeV for the current case. 

Fig.~\ref{fig:nh2_eta} shows allowed regions for muon $g-2$ in terms of the mass of DM. 
As can be seen the figure, the muon $g-2$, $\Delta a_\mu \lesssim 3.0 \times 10^{-18}$, has almost no discrepancy from the SM prediction, and the predictions tend to be smaller than that of the $S_0$ DM case, due to the large mass of $S_0$. 
Different from the $S_0$ DM case, $m_S$ mass runs whole the range within our fixed range. 

\subsubsection{\rm IH}

\begin{figure}[tb]
\begin{center}
\includegraphics[width=53mm]{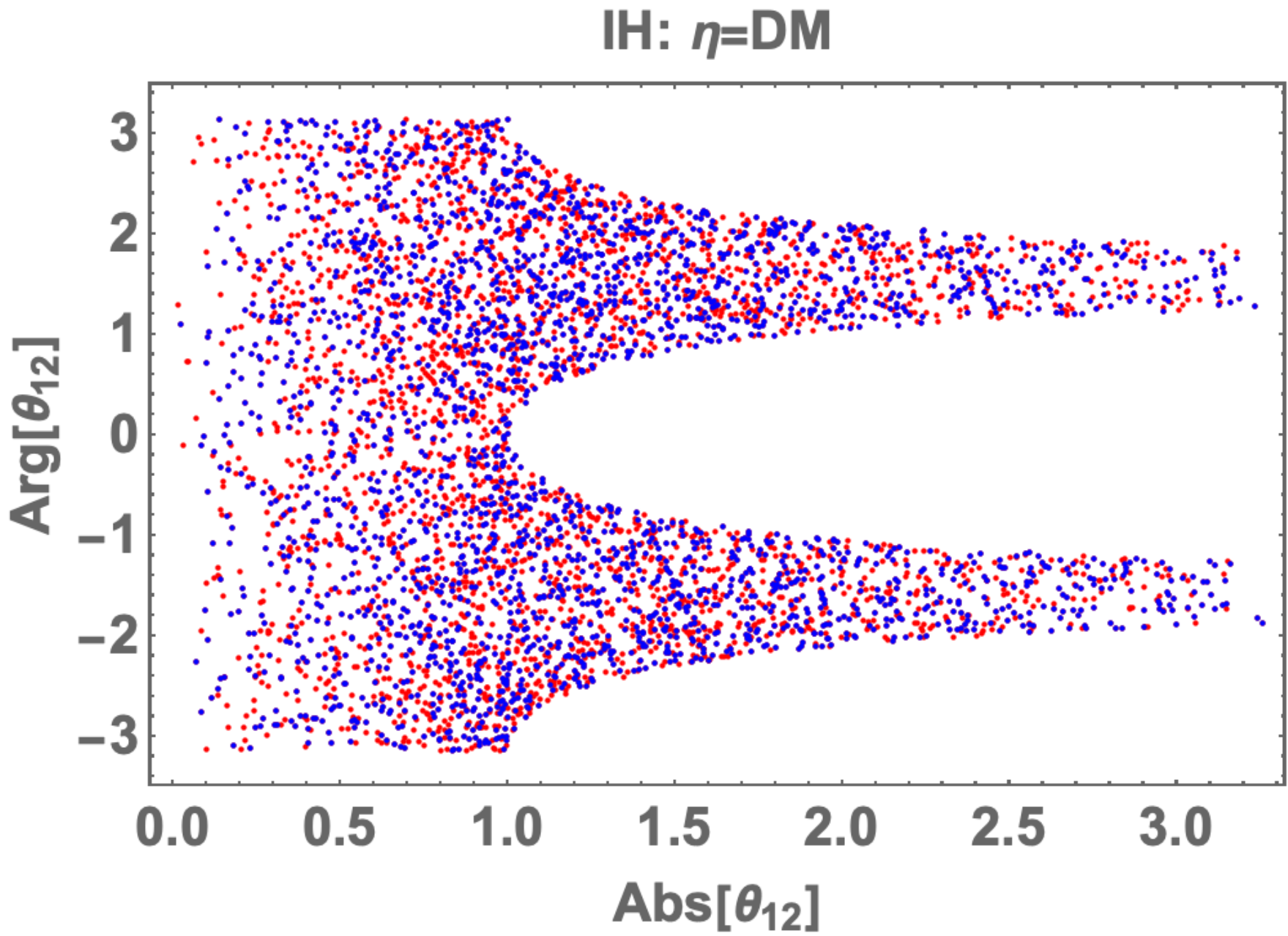}
\includegraphics[width=53mm]{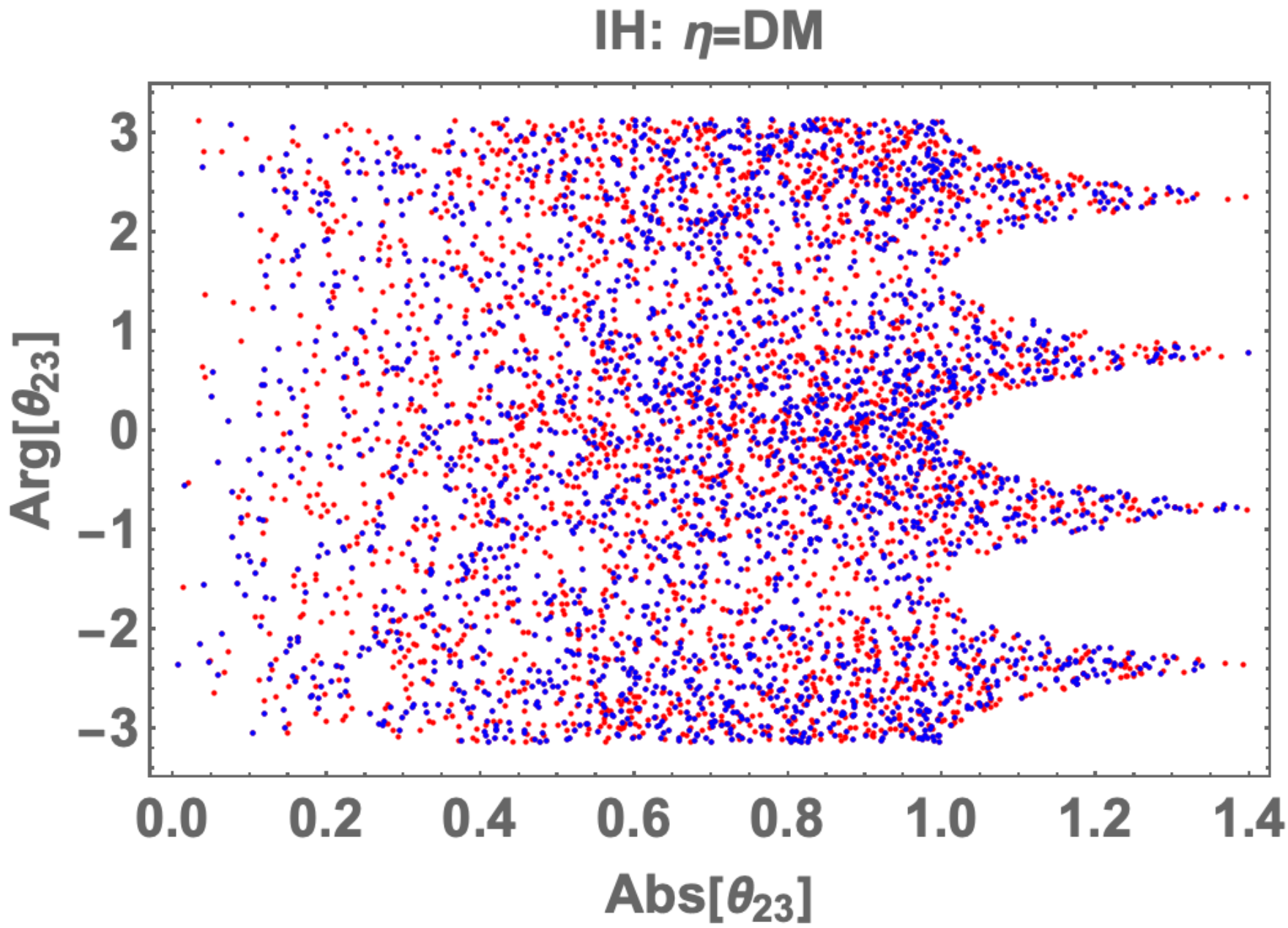}
\includegraphics[width=53mm]{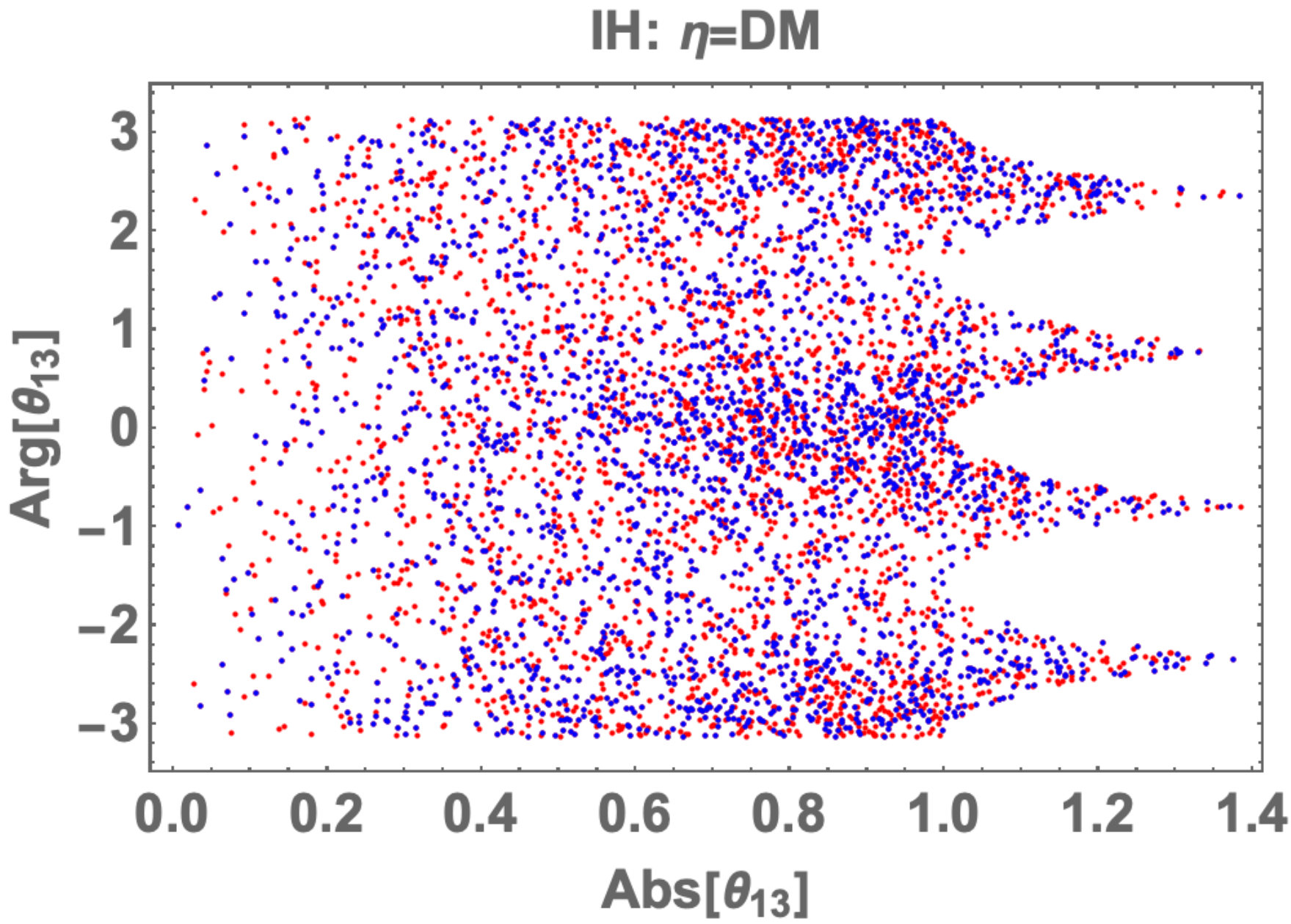}
\caption{Same plot as Fig.~\ref{fig:nh1_eta}, for the IH case with $\eta$ DM. }
\label{fig:ih1_eta}
\end{center}
\end{figure}
\begin{figure}[tb]
\begin{center}
\includegraphics[width=88mm]{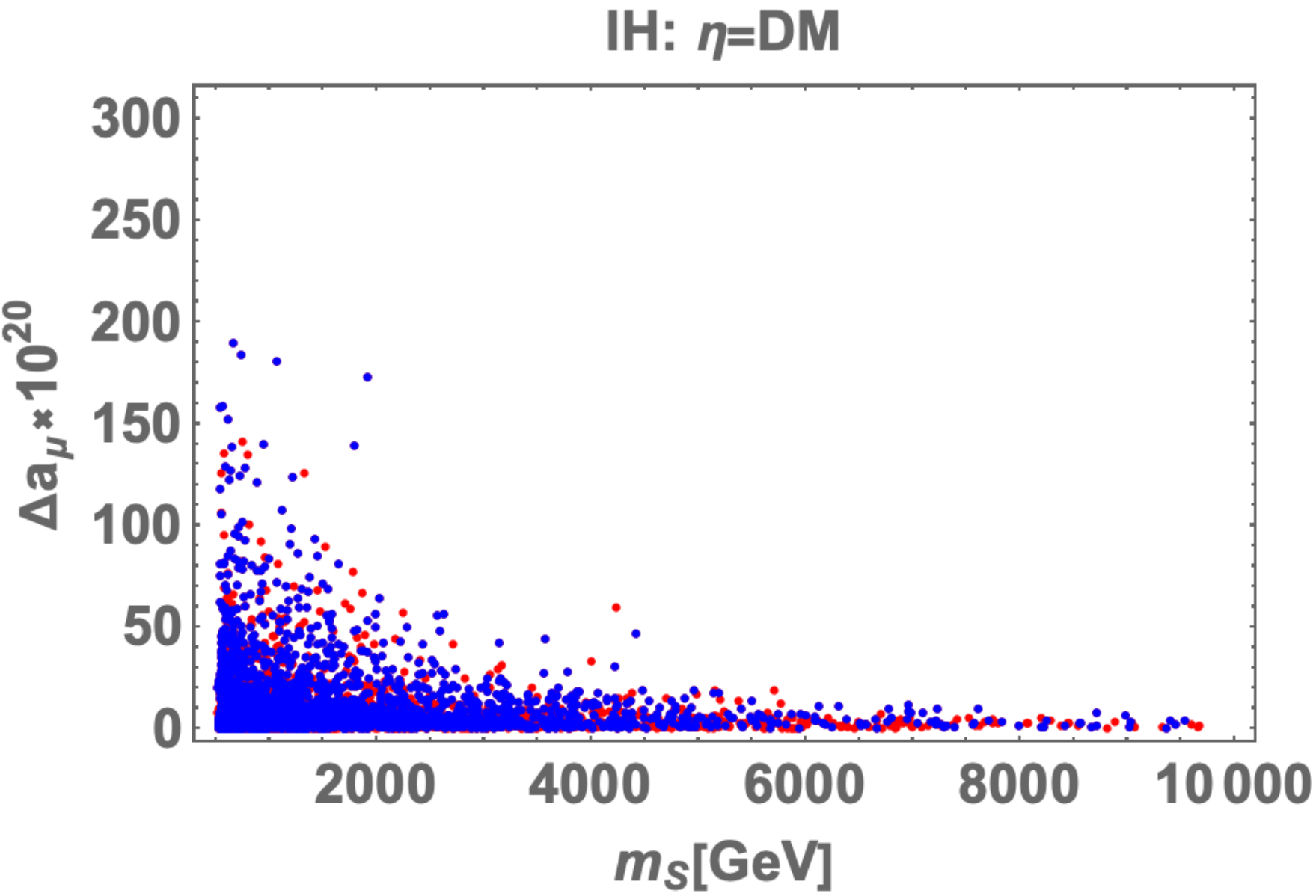}
\caption{Same plot as Fig.~\ref{fig:nh2_eta}, for the IH case with $\eta$ DM. }
\label{fig:ih3}
\end{center}
\end{figure}
Fig.~\ref{fig:ih1_eta} shows allowed regions for $\theta_{12}$ (left), $\theta_{23}$ (center), $\theta_{13}$ (right). 
Similar to the NH case, these figures imply that these shapes are almost the same as the NH one but more points are out of $\sum D_\nu \le 120$ meV, as we observed in the $S_0$ DM case. 
However, the number of blue points is larger than that of the IH case with $S_0$ DM. 

Fig.~\ref{fig:ih3} shows allowed regions for the muon $g-2$ in terms of the mass of DM. 
Similar to the NH case, the muon $g-2$ has almost no discrepancy from the SM prediction; $\Delta a_\mu \lesssim 2.0 \times 10^{-18}$. 
Again, $m_S$ mass runs whole the range within our fixed range, and the similar conclusion in the NH case for the muon $g-2$ prediction can be seen in the figure.

\section{Summary and discussion}
\label{sec:summary}

We have proposed a radiatively induced neutrino masses at the three-loop level, based on the Ma model. 
We introduced a non-invertible symmetry in the class under the ${\mathbb Z_2}$ gauging of ${\mathbb Z_6}$ symmetry and added three isospin doublet vector-like fermions $L'$ and singlet boson $S_0$. 
The Yukawa interactions directly related to the neutrino masses are forbidden at the tree-level, under this symmetry. 
However, the terms are allowed at one-loop level due to $L'$ and $S_0$ as well as $\eta$, which is no longer invariant under this symmetry any more. 
Therefore, the symmetry is dynamically broken. Interestingly, $\eta$ contributes to both the radiative matrices $y^\eta$ and $m_\nu$. 

Under the model set, we have performed numerical analyses to satisfy the lepton flavor violations, muon $g-2$, and boson DM candidates $S_0$ and $\eta_R$ for the cases of NH and IH. 
Then, we have shown allowed parameter space for our model in four cases and concluded all the cases have almost same results except the DM candidate. 
Although there are similarity in the results, we have found specific features for each DM case, especially in the results of Abs$[\theta_{23, 13}] > 1.0$.

\section*{Acknowledgments}
HO is supported by Zhongyuan Talent (Talent Recruitment Series) Foreign Experts Project. 
YS is supported by Natural Science Foundation of China under grant No. W2433006. 

\bibliography{ctma4}
\end{document}